\documentclass[sigconf]{acmart}

\usepackage{booktabs} 
\usepackage{epsfig}
\usepackage{amsmath}
\usepackage{amssymb}
\usepackage{helvet}
\usepackage{courier}
\usepackage{multirow}
\usepackage{amsopn}
\usepackage{graphicx,subfigure}
\usepackage{array}
\usepackage{xcolor}
\usepackage{url}
\usepackage{setspace}
\usepackage{bm}
\usepackage{tabularx}
\usepackage{algorithm}
\usepackage{epstopdf}
\usepackage{tikz}
\usetikzlibrary{fit,positioning}
\definecolor{mygray}{gray}{.9}
\definecolor{mypink}{rgb}{.99,.91,.95}
\definecolor{mycyan}{cmyk}{.3,0,0,0}

\copyrightyear{2018}
\acmYear{2018}
\setcopyright{acmcopyright}
\acmConference[MM'18]{2018 ACM Multimedia Conference}{}{Seoul, Korea}
\acmBooktitle{2018 ACM Multimedia Conference (MM'18), October, 2018, Seoul, Korea}
\acmPrice{15.00}
\acmDOI{10.1145/3240508.3240528}
\acmISBN{978-1-4503-5665-7/18/10}

\begin{document}
\title{Semi-supervised Deep Generative Modelling of Incomplete Multi-Modality Emotional Data}

\author{Changde Du}
\affiliation{%
  \institution{Research Center for Brain-inspired Intelligence \& National Laboratory of Pattern Recognition, CASIA}
  \institution{University of CAS, Beijing, China}
}
\email{duchangde@gmail.com}

\author{Changying Du, Hao Wang}
\orcid{1234-5678-9012}
\affiliation{%
  \institution{360 Search Lab}
  \city{Beijing}
  \state{China}
}
\email{ducyatict@gmail.com}
\email{cashenry@126.com}

\author{Jinpeng Li}
\affiliation{%
  \institution{Research Center for Brain-inspired Intelligence \& National Laboratory of Pattern Recognition, CASIA}
  \institution{University of CAS, Beijing, China}
  }
\email{lijinpeng2015@ia.ac.cn}

\author{Wei-Long Zheng}
\affiliation{%
  \institution{Department of Computer Science and Engineering, SJTU}
  \city{Shanghai}
  \state{China}
}
\email{weilong@sjtu.edu.cn}

\author{Bao-Liang Lu}
\affiliation{%
  \institution{Department of Computer Science and Engineering, SJTU}
  \city{Shanghai}
  \state{China}
}
\email{bllu@sjtu.edu.cn}

\author{Huiguang He}
\authornote{Dr. Huiguang He is the corresponding author and he is also with the Center for Excellence in Brain Science and Intelligence
Technology, Chinese Academy of Sciences and the
University of Chinese Academy of Sciences.
}
\affiliation{%
  \institution{Research Center for Brain-inspired Intelligence \& National Laboratory of Pattern Recognition, CASIA}
}
\email{huiguang.he@ia.ac.cn}

\renewcommand{\shortauthors}{Du et al.}

\begin{abstract}
There are threefold challenges in emotion recognition. First, it is difficult to recognize human's emotional states only considering a single modality.  Second, it is expensive to manually annotate the emotional data. Third, emotional data often suffers from missing modalities due to unforeseeable sensor malfunction or configuration issues.  In this paper, we address all these problems under a novel multi-view deep generative framework. Specifically, we propose to model the statistical relationships of multi-modality emotional data using multiple modality-specific generative networks with a shared latent space. By imposing a Gaussian mixture assumption on the posterior approximation of the shared latent variables, our framework can learn the joint deep representation from multiple modalities and evaluate the importance of each modality simultaneously. To solve the labeled-data-scarcity problem, we extend our multi-view model to semi-supervised learning scenario by casting the semi-supervised classification problem as a specialized missing data imputation task.  To address the missing-modality problem, we further extend our semi-supervised multi-view model to deal with incomplete data, where a missing view is treated as a latent variable and integrated out during inference. This way, the proposed overall framework can utilize all available (both labeled and unlabeled, as well as both complete and incomplete) data to improve its generalization ability. The experiments conducted on two real multi-modal emotion datasets demonstrated the superiority of our framework.
\end{abstract}


\maketitle

\section{Introduction}
With the development of human-computer interaction (HCI), emotion recognition has become increasingly important. Since human's emotion contains many nonverbal cues, various modalities ranging from facial expressions, body gesture, voice to physiological signals can be used as the indicators of emotional states ~\cite{calvo2010affect,poria2017review}.
In real-world applications, it is difficult to recognize human's emotional states only considering a single modality, because signals from different modalities represent different aspects of emotion and provide complementary information.  Recent studies show that integrating multiple modalities can significantly boost the emotion recognition accuracy ~\cite{lu2015combining,Ranganathan2016Multimodal,soleymani2016analysis}.
The most successful approach to fuse the information from multiple modalities is based on deep multi-view representation learning ~\cite{ngiam2011multimodal,srivastava2014multimodal,wang2015deep}. E.g., ~\cite{pang2015deep} proposed a joint density model for emotion analysis with a multi-modal deep Boltzmann machine (DBM) ~\cite{srivastava2014multimodal}. This multi-modal DBM is exploited to model the joint distribution over visual, auditory, and textual features.
~\cite{Liu2016Emotion} proposed a multi-modal emotion recognition method by using multi-modal autoencoders (MAE) ~\cite{ngiam2011multimodal}, in which the joint representations of Electroencephalogram (EEG) and eye movement signals were extracted.
Nevertheless, there are still limitations with these deep multi-modal emotion recognition methods, e.g., their performances depend on the amount of labeled data and they could not handle incomplete data.

By using the modern sensing equipments, we can easily collect massive emotion-related data from multiple modalities. But, the data labeling procedure requires lots of manual efforts. Therefore, in most cases only a small set of labeled samples is available, while the majority of whole dataset is left unlabeled.  In addition to challenges with insufficient labeled data, one must often address the incomplete-data problem, i.e., not all modalities are available for every data point. Generally, we can identify various causes for incomplete data. E.g., unforeseeable sensor malfunction may fail to collect sensing information, thus providing us incomplete data with one or more missing modalities. Traditional multi-modal emotion recognition approaches ~\cite{pang2015deep,lu2015combining,Liu2016Emotion} only utilized the limited amount of labeled data, which may result in severe overfitting.  Also, most of them neglect the missing modality issue, which greatly limits their applications in real-world scenarios.  The most attractive way to deal with the aforementioned issues is semi-supervised learning (SSL) with incomplete data. SSL can improve model's generalization ability by exploiting both labeled and unlabeled data simultaneously ~\cite{schels2014using,jia2014novel,zhang2016enhanced}, and learning from incomplete data can guarantee the robustness of the emotion recognition system ~\cite{wagner2011exploring}.

In this paper, we show that the problems mentioned above can be resolved under a unified multi-view deep generative framework. For modeling the statistical relationships of multi-modality emotional data, a shared latent variable is transformed by different modality-specific generative networks to different data views (modalities).
Instead of treating each view equally, we impose a non-uniformly weighted Gaussian mixture assumption on the posterior approximation of the shared latent variables.  This is critical for inferring the joint latent representation and the weight factor of each view from multiple modalities.  During optimization, a second lower bound to the variational lower bound is derived to address the intractable entropy of a mixed Gaussians.
To leverage the contextual information in the unlabeled data to augment the limited labeled data, we then extend our multi-view framework to SSL scenario. It is achieved by casting the semi-supervised classification problem as a specialized missing data imputation task.  Specifically, we treat the unknown labels as latent variables and estimate them within a multi-view auto-encoding variational Bayes framework.
We further extend the proposed SSL algorithm to the incomplete-data case by introducing latent variables for the missing views. Besides the unknown labels, the missing views are also integrated out so that the marginal likelihood is maximized with respect to model parameters. In this way, our SSL algorithm can utilize all available data: both labeled and unlabeled, as well as both complete and incomplete. Since the category information and the uncertainty of missing view are taken into account in the training process, our SSL algorithm is more powerful than traditional missing view imputation methods ~\cite{Quanz2012CoNet,wang2015deep,Hastie2015Matrix,Chandar2016Correlational}. We finally demonstrate the superiority of our framework and provide insightful observations on two real multi-modal emotion datasets.
\vspace{-0.05cm}
\section{Related Work}
Multi-modal approaches have been widely implemented for emotion recognition ~\cite{pang2015deep,lu2015combining,soleymani2016analysis,Liu2016Emotion,Ranganathan2016Multimodal,Tzirakis2017End,Zheng2018EmotionMeter}. E.g., ~\cite{Ranganathan2016Multimodal} used a multi-modal deep belief network (DBN) to extract features from face, body gesture, voice and physiological signals for emotion classification. ~\cite{lu2015combining} classified the combination of EEG and eye movement signals into three affective states. But, very few of them explored SSL. To the best of our knowledge, only ~\cite{zhang2016enhanced} proposed an enhanced multi-modal co-training algorithm for semi-supervised emotion recognition, but its shallow structure is hard to capture the high-level correlation between different modalities. In addition, most prior work in this field assumes that all modalities are available at all times ~\cite{zhang2016enhanced,Zheng2018EmotionMeter}, which is not realistic in practical environments. In contrast to the above methods, our framework naturally allows us to perform multi-modal emotion recognition within SSL and incomplete-data situations.

The variational autoencoder (VAE) ~\cite{VAE,rezende2014stochastic} is one of the most popular deep generative models (DGMs). VAE has shown great advantages in semi-supervised classification ~\cite{kingma2014semi,maaloe2016auxiliary}. E.g., Kingma et al. ~\cite{kingma2014semi} proposed a semi-supervised VAE (M2) by modeling the joint distribution over data and labels. ~\citeauthor{maaloe2016auxiliary} proposed the auxiliary DGMs (ADGM and SDGM) ~\cite{maaloe2016auxiliary} by introducing auxiliary variables, which improve the variational approximation.
However, these models cannot effectively deal with multi-view data, especially in incomplete-view case.
Our proposed semi-supervised multi-view DGMs distinguish our method from all existing ones using VAE framework ~\cite{Wang2016Deep,Burda2016Importance,Kingma2016Improving,serban2016multi,maaloe2016auxiliary}.

Incomplete-data problem is often circumvented via imputation methods ~\cite{Williams2007On,Amini2009Learning,Wang2010Classification,Xu2015Multi,Zhang2018Multi}. Common imputation schemes include matrix completion ~\cite{Keshavan2009Matrix,Hazan2015Classification,Hastie2015Matrix} and autoencoder-based methods ~\cite{wang2015deep,Chandar2016Correlational,Luan2017Missing,Shang2017VIGAN}. Matrix completion methods, such as SoftImputeALS ~\cite{Hastie2015Matrix}, focus on imputing the missing entries of a partially observed matrix based on assumption that the completed matrix has a low-rank structure. Matrix completion methods often assume data is missing at random (MAR), which might not be optimal for our problem where modalities are missing at continuous blocks. On the other hand, autoencoder-based methods, such as DCCAE ~\cite{wang2015deep} and CorrNet ~\cite{Chandar2016Correlational}, exploit the connections between views, enabling the incomplete view to be restored with the help of the complete view.
Besides low-rank structure of the data matrix and the connections between views, category information is also important for missing view imputation tasks, though category labels may be partially observed. So far, very few algorithms ~\cite{Yu2011Bayesian,Quanz2012CoNet} can estimate the missing view under the SSL scenario. Although CoNet ~\cite{Quanz2012CoNet} utilized deep neural networks (DNNs) to predict the missing view based on existing views and partially observed labels, its feedforward structure could not integrate multiple views effectively in classification. Additionally, most previous works treat the missing data as fixed values and hence ignore the uncertainty of the missing data. Unlike them, our SiMVAE essentially performs infinite imputations by integrating out the missing data.

\section{Methodology}
In this section, we first develop a multi-view variational autoencoder (MVAE) model for fusing multi-modality emotional data. Based on MVAE, we further build a semi-supervised emotion recognition algorithm. Finally, we develop a more robust semi-supervised algorithm to address the incomplete multi-modality emotional data.
For simplicity we restrict further discussion to the case of two views, though all the proposed methods can be extended to
more than two views. Assume we are faced with multi-view data that appears as pairs $(\mathfrak{X},\  y) = (\{\mathrm{\mathbf{x}}^{(v)}\}_{v=1}^{2},\  y)$, with  observation $\mathrm{\mathbf{x}}^{(v)}$ from the $v$-th view and the corresponding class label $y$.
\vspace{-0.05cm}
\subsection{Multi-view Variational Autoencoder}
\subsubsection{DNN-parameterized Likelihoods}
\begin{figure*}[t]
\centering
  \subfigure[$P_1$] {
            \begin{tikzpicture}[line width = 0pt]
            \tikzstyle{main}=[circle, thick, draw =black!100, node distance = 6mm]
            \tikzstyle{connect}=[-latex, thick]
            \tikzstyle{box}=[rectangle, draw=black!100]
              \node[main, scale=0.8, minimum size = 8mm] (z) [label=above:\footnotesize  $p(\mathrm{\mathbf{z}})$] {$\mathrm{\mathbf{z}}$ };
              \node[main, scale=0.68, minimum size = 8mm, fill = black!12] (x_1) [below=of z,xshift=-1.0cm, yshift=0.1cm, label=below:] {$\mathrm{\mathbf{x}}^{(1)}$};
              \node[main, scale=0.68, minimum size = 8mm, fill = black!12] (x_V) [below=of z,xshift=1.0cm, yshift=0.1cm, label=below: \scriptsize {$\hspace{-1.3cm} \{p_{\theta^{(v)}}(\mathrm{\mathbf{x}}^{(v)}|\mathrm{\mathbf{z}})\}_{v=1}^2$}] {\small $\mathrm{\mathbf{x}}^{(2)}$ };
              \path (z) edge [connect] (x_1)
                    (z) edge [connect] (x_V);
            \coordinate (ne)at(current bounding box.north east);
            \coordinate (sw)at(current bounding box.south west);
            \end{tikzpicture}}
  \hspace {0.0cm}
  \subfigure [$Q_1$] {
            \begin{tikzpicture}[line width = 0pt]
            \tikzstyle{main}=[circle,  thick, draw =black!100, node distance = 6mm]
            \tikzstyle{connect}=[-latex, thick]
            \tikzstyle{box}=[rectangle, draw=black!100]
            \useasboundingbox(ne)rectangle(sw);
              \node[main, scale=0.8, minimum size = 8mm] (z) [label=above: \footnotesize {$q_\phi(\mathrm{\mathbf{z}}|\mathfrak{X})$}] {$\mathrm{\mathbf{z}}$ };
              \node[main, scale=0.68, minimum size = 8mm, fill = black!12] (x_1) [below=of z,xshift=-1.1cm, yshift=0.1cm, label=below:\footnotesize] {$\mathrm{\mathbf{x}}^{(1)}$};
              \node[main, scale=0.68, minimum size = 8mm, fill = black!12] (x_V) [below=of z,xshift=1.1cm, yshift=0.1cm, label=below:\footnotesize] {\small$\mathrm{\mathbf{x}}^{(2)}$ };
              \path (x_1) edge [connect] (z)
                    (x_V) edge [connect] (z);
            \end{tikzpicture}}
  \hspace {0.3cm}
  \subfigure[$P_2$] {
            \begin{tikzpicture}[line width = 0pt]
            \tikzstyle{main}=[circle, thick, draw =black!100, node distance = 5mm]
            \tikzstyle{connect}=[-latex, thick]
            \tikzstyle{box}=[rectangle, draw=black!100]
              \draw[fill = black!12] (-0.0,-0.31) arc (-90:-270:0.32cm);
              \node[main, scale=0.8, minimum size = 8mm] (y) [label=above:\footnotesize  $p(y)$] { $y$};
              \node[main, scale=0.8, minimum size = 8mm] (z) [right=of y, xshift=0.2cm,label=above:\footnotesize  $p(\mathrm{\mathbf{z}})$] {$\mathrm{\mathbf{z}}$ };
              \node[main, scale=0.68, minimum size = 8mm, fill = black!12] (x_1) [below=of y,label=below:] {$\mathrm{\mathbf{x}}^{(1)}$};
              \node[main, scale=0.68, minimum size = 8mm, fill = black!12] (x_V) [below=of z,label=below:\scriptsize {$\hspace{-1.0cm} \{ p_{\theta^{(v)}}(\mathrm{\mathbf{x}}^{(v)}|y, \mathrm{\mathbf{z}})\}_{v=1}^2$}] {\small $\mathrm{\mathbf{x}}^{(2)}$ };
              \path (y) edge [connect] (x_1)
                    (z) edge [connect] (x_1)
		            (y) edge [connect] (x_V)
                    (z) edge [connect] (x_V);
            \coordinate (ne)at(current bounding box.north east);
            \coordinate (sw)at(current bounding box.south west);
            \end{tikzpicture}}
  \hspace {0.1cm}
  \subfigure [$Q_2$] {
            \begin{tikzpicture}[line width = 0pt]
            \tikzstyle{main}=[circle, thick, draw =black!100, node distance = 5mm]
            \tikzstyle{connect}=[-latex, thick]
            \tikzstyle{box}=[rectangle, draw=black!100]
            \useasboundingbox(ne)rectangle(sw);
              \draw[fill = black!12] (-0.0,-0.31) arc (-90:-270:0.32cm);
              \node[main, scale=0.8, minimum size = 8mm] (y) [label=above:\footnotesize $q_{\varphi}(y|\mathfrak{X})$] {$y$};
              \node[main, scale=0.8, minimum size = 8mm] (z) [right=of y, xshift=0.2cm,label=above: \footnotesize {$q_\phi(\mathrm{\mathbf{z}}|\mathfrak{X}, y)$}] {$\mathrm{\mathbf{z}}$ };
              \node[main, scale=0.68, minimum size = 8mm, fill = black!12] (x_1) [below=of y,label=below:\footnotesize] {$\mathrm{\mathbf{x}}^{(1)}$};
              \node[main, scale=0.68, minimum size = 8mm, fill = black!12] (x_V) [below=of z,label=below:\footnotesize] {\small$\mathrm{\mathbf{x}}^{(2)}$ };
              \path (x_1) edge [connect] (y)
                    (x_1) edge [connect] (z)
		            (x_V) edge [connect] (y)
                    (x_V) edge [connect] (z)
                    (y)   edge [connect] (z);
            \end{tikzpicture}}
  \hspace {-0.5cm}
  \subfigure[$P_3$] {
            \begin{tikzpicture}[line width = 0pt]
            \tikzstyle{main}=[circle, thick, draw =black!100, node distance = 6mm]
            \tikzstyle{connect}=[-latex, thick]
            \tikzstyle{box}=[rectangle, draw=black!100]
              \draw[fill = black!12] (-0.0,-0.31) arc (-90:-270:0.32cm);
              \draw[fill = black!12] (1.42,-1.58) arc (-90:-270:0.32cm);
              \node[main, scale=0.8, minimum size = 8mm] (y) [label=above:\footnotesize  $p(y)$] { $y$};
              \node[main, scale=0.8, minimum size = 8mm] (z) [right=of y, xshift=0.2cm,label=above:\footnotesize  $p(\mathrm{\mathbf{z}})$] {$\mathrm{\mathbf{z}}$ };
              \node[main,  scale=0.8, minimum size = 8mm,  fill = black!12] (x_1) [below=of y, label=below:\scriptsize {$p_{\theta^{o}}(\mathrm{\mathbf{x}}^{o}|y, \mathrm{\mathbf{z}}, \mathrm{\mathbf{x}}^{m})$}] {$\mathrm{\mathbf{x}}^{o}$};
              \node[main,  scale=0.8, minimum size = 8mm] (x_2) [below=of z,label=below:\scriptsize {$\qquad \quad p_{\theta^{m}}(\mathrm{\mathbf{x}}^{m}|y, \mathrm{\mathbf{z}})$}] {\small $\mathrm{\mathbf{x}}^{m}$ };
              \path (y) edge [connect] (x_1)
                    (z) edge [connect] (x_1)
		            (y) edge [connect] (x_2)
                    (x_2) edge [connect] (x_1)
                    (z) edge [connect] (x_2);
            \coordinate (ne)at(current bounding box.north east);
            \coordinate (sw)at(current bounding box.south west);
            \end{tikzpicture}}
  \hspace {-1.2cm}
  \subfigure [$Q_3$] {
            \begin{tikzpicture}[line width = 0pt]
            \tikzstyle{main}=[circle, thick, draw =black!100, node distance = 6mm]
            \tikzstyle{connect}=[-latex, thick]
            \tikzstyle{box}=[rectangle, draw=black!100]
            \useasboundingbox(ne)rectangle(sw);
              \draw[fill = black!12] (-0.0,-0.31) arc (-90:-270:0.32cm);
              \draw[fill = black!12] (1.42,-1.58) arc (-90:-270:0.32cm);
              \node[main, scale=0.8, minimum size = 8mm] (y) [label=above:\footnotesize $q_{\varphi}(y|\mathfrak{X})$] {$y$};
              \node[main, scale=0.8, minimum size = 8mm] (z) [right=of y, xshift=0.2cm,label=above: \footnotesize {$q_\phi(\mathrm{\mathbf{z}}|\mathfrak{X}, y)$}] {$\mathrm{\mathbf{z}}$ };
              \node[main, scale=0.8, minimum size = 8mm, fill = black!12] (x_1) [below=of y,label=below:\footnotesize] {$\mathrm{\mathbf{x}}^{o}$};
              \node[main, scale=0.8, minimum size = 8mm] (x_2) [below=of z,label=below:\footnotesize {$q_{\psi}(\mathrm{\mathbf{x}}^{m}| \mathrm{\mathbf{x}}^{o})$}] {\small$\mathrm{\mathbf{x}}^{m}$ };
              \path (x_1) edge [connect] (y)
                    (x_1) edge [connect] (z)
		            (x_2) edge [connect] (y)
                    (x_2) edge [connect] (z)
                    (x_1) edge [connect] (x_2)
                    (y)   edge [connect] (z);
            \end{tikzpicture}}
\vspace {-0.5cm}
\caption{Graphical models of the proposed algorithms: (a, b) multi-view variational autoencoder (MVAE); (c, d) semi-supervised MVAE (SMVAE); (e, f) semi-supervised incomplete MVAE (SiMVAE). In (e) and (f), we partition the two-view data point (i.e., $\mathfrak{X} = \{\mathrm{\mathbf{x}}^{(1)}, \mathrm{\mathbf{x}}^{(2)}\}$) into an observed view $\mathrm{\mathbf{x}}^{o}$ and a missing view $\mathrm{\mathbf{x}}^{m}$ (i.e., $\mathfrak{X} = \{\mathrm{\mathbf{x}}^{o}, \mathrm{\mathbf{x}}^{m}\}$). Both $y$ and $\mathrm{\mathbf{x}}^{m}$ are partially observed.}
\label{fig:MVAE}
\vspace {-0.3cm}
\end{figure*}
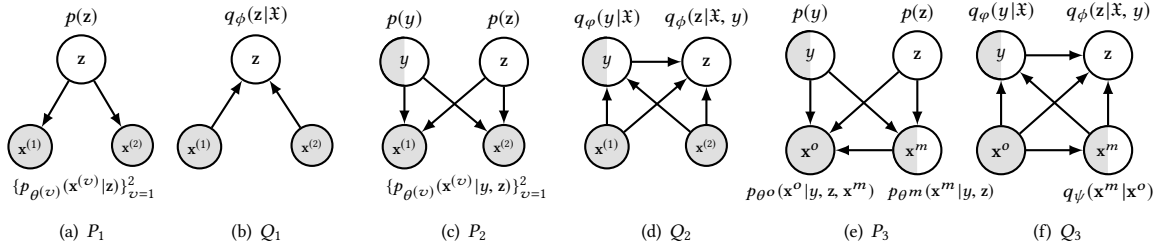

We assume that multiple data views (modalities) $\{\mathrm{\mathbf{x}}^{(v)}\}_{v=1}^{2}$ are generated independently from a shared latent space with  multiple view-specific generative networks.
Specifically, we assume a shared latent variable $\mathrm{\mathbf{z}}$ generates $\mathrm{\mathbf{x}}^{(v)}$ with the following generative model $P_1$ (cf. Figure \ref{fig:MVAE}a):
\begin{align}
 p_{\theta^{(v)}}(\mathrm{\mathbf{x}}^{(v)}|\mathrm{\mathbf{z}})  &= f(\mathrm{\mathbf{x}}^{(v)}; \mathrm{\mathbf{z}}, \theta^{(v)}), \quad v\in\{1, 2\},
\end{align}
where $f(\mathrm{\mathbf{x}}^{(v)}; \mathrm{\mathbf{z}}, \theta^{(v)})$ is a suitable likelihood function (e.g. a Gaussian for continuous observation or Bernoulli for binary observation), which is formed by a non-linear transformation of the latent variable $\mathbf{z}$. This non-linear transformation is essential to allow for higher moments of the data to be captured by the density model, and we choose these non-linear functions to be DNNs, referred to as the generative networks, with parameters $\{\theta^{(v)}\}_{v=1}^{2}$. Note that, the likelihoods for different views are assumed to be independent of each other, with  potentially different DNN types for different modalities.

\subsubsection{Gaussian Prior and Gaussian Mixture Posterior}
In vanilla VAE ~\cite{VAE,rezende2014stochastic}, which can only handle single-view data, both the prior $p(\mathrm{\mathbf{z}})$ and the approximate posterior  $q_{\phi}(\mathrm{\mathbf{z}}|\mathfrak{X})$ are assumed to be Gaussian distributions in order to maintain mathematical and computational tractability. Although this assumption has leaded to favorable results on several tasks, it is clearly a restrictive and often unrealistic assumption. Specifically, the choice of a Gaussian distribution for $p(\mathrm{\mathbf{z}})$ and $q_{\phi}(\mathrm{\mathbf{z}}|\mathfrak{X})$ imposes a strong uni-modal structure assumption on the latent space. However, for data distributions that are strongly multi-modal, the uni-modal Gaussian assumption inhibits the model's ability to extract and represent important structure in the data.
To improve the flexibility of the model, one way is to impose a Mixture of Gaussians (MoG) assumption on $p(\mathrm{\mathbf{z}})$. However, it has the risk of creating separate ``islands" of discontinuous manifolds that may break the meaningfulness of the representation in the latent space.

To learn more powerful and expressive models (in particular, models with multi-modal latent variable structures for multi-modal emotion recognition
applications) we seek a MoG for $q_{\phi}(\mathrm{\mathbf{z}}|\mathfrak{X})$, while preserving $p(\mathrm{\mathbf{z}})$ as a standard Gaussian. Thus the prior distribution and the inference model $Q_1$ (cf. Figure \ref{fig:MVAE}b) are defined as: $p(\mathrm{\mathbf{z}}) = \mathcal{N}\left(\mathrm{\mathbf{z}}|\bm{0}, \mathrm{\mathbf{I}}\right)$,
\begin{align}\label{Posterior}
q_\phi(\mathrm{\mathbf{z}}|\mathfrak{X})  & = \sum\limits_{v=1}^{2}\lambda^{(v)}\mathcal{N}\left(\mathrm{\mathbf{z}}|\bm{\mu}_{\phi^{(v)}}(\mathrm{\mathbf{x}}^{(v)}),\ \bm{\Sigma}_{\phi^{(v)}}(\mathrm{\mathbf{x}}^{(v)}) \right),
\end{align}
where the mean $\bm{\mu}_{\phi^{(v)}}$ and the covariance $\bm{\Sigma}_{\phi^{(v)}}$ are nonlinear functions of the observation $\mathrm{\mathbf{x}}^{(v)}$, with variational parameter $\bm{\phi}^{(v)}$. As in our generative model, we choose these nonlinear functions to be DNNs, referred to as the inference networks. $\lambda^{(v)}$ is the non-negative normalized weight factor for the $v$-th view, i.e., $\lambda^{(v)} > 0$ and $\sum_{v=1}^2\lambda^{(v)} =1$. Note that, Gershman et al. \cite{gershman2012nonparametric} proposed a nonparametric variational inference method by simply assuming the variational distribution to be a uniformly weighted Gaussian mixture. However, treating each component equally will lose flexibility in fusing multiple data views. Instead of treating each view equally, our non-uniformly weighted Gaussian mixture assumption can weight each view automatically in subsequent emotion recognition tasks, which is useful to identify the importance of each view.

\subsection{Semi-supervised Multi-modal Emotion Recognition}
Although many supervised emotion recognition algorithms exist (see ~\cite{poria2017review} for a thorough literature review), very few
semi-supervised algorithms have been proposed to improve the recognition performance by utilizing both labeled and unlabeled data.
Here we extend MVAE by introducing a conditional probabilistic distribution for the unknown labels to obtain a semi-supervised multi-view classification algorithm.

\subsubsection{Generative model $P_2$}
Since the emotional data is continuous, we choose the Gaussian likelihoods.
Then our generative model $P_2$ (cf. Figure \ref{fig:MVAE}c) is defined as $p(y)$$p(\mathrm{\mathbf{z}})$$\prod_{v=1}^{2}$$p_{\theta^{(v)}}(\mathrm{\mathbf{x}}^{(v)}|y, \mathrm{\mathbf{z}})$:
\begin{align} \label{SMVAE_generative}
p(y) &= \mathrm{Cat}\left(y|\bm{\pi}\right), \qquad p(\mathrm{\mathbf{z}}) = \mathcal{N}\left(\mathrm{\mathbf{z}}|\bm{0}, \mathrm{\mathbf{I}}\right),\\
\nonumber p_{\theta^{(v)}}(\mathrm{\mathbf{x}}^{(v)}|y, \mathrm{\mathbf{z}}) & = \mathcal{N}\left(\bm{\mu}_{\theta^{(v)}}(y, \mathbf{z}),\ \mathrm{diag}(\bm{\sigma}^{2}_{\theta^{(v)}}(y, \mathbf{z})) \right),
\end{align}
where $\mathrm{Cat}(\cdot)$ denotes the categorical distribution, $y$ is treated as a latent variable for the unlabeled data points,
and the mean $\bm{\mu}_{\theta^{(v)}}$ and variance $\bm{\sigma}^{2}_{\theta^{(v)}}$ are nonlinear functions of $y$ and $\mathbf{z}$, with parameter $\theta^{(v)}$.

\subsubsection{Inference model $Q_2$}
The inference model $Q_2$ (cf. Figure \ref{fig:MVAE}d) is defined as $q_{\varphi}(y|\mathfrak{X})q_\phi(\mathrm{\mathbf{z}}|\mathfrak{X}, y)$:
\begin{align} \label{SMVAE_inference}
 q_{\varphi}(y|\mathfrak{X}) &= \mathrm{Cat}\left(y|\bm{\pi}_{\varphi}(\mathfrak{X})\right),\\
\nonumber q_\phi(\mathrm{\mathbf{z}}|\mathfrak{X}, y)  &= \sum\limits_{v=1}^{2}\lambda^{(v)}\mathcal{N}\left(\mathrm{\mathbf{z}}|\bm{\mu}_{\phi^{(v)}}(\mathrm{\mathbf{x}}^{(v)}, y),\ \bm{\Sigma}_{\phi^{(v)}}(\mathrm{\mathbf{x}}^{(v)}, y) \right),
\end{align}
where $q_{\varphi}(y|\mathfrak{X})$ is the introduced conditional distribution for $y$, and $q_\phi(\mathrm{\mathbf{z}}|\mathfrak{X}, y)$ is assumed to be a mixture of Gaussians to combine the information from multiple data views and the label. Intuitively,  $q_\phi(\mathrm{\mathbf{z}}|\mathfrak{X}, y)$, $p_{\theta^{(v)}}(\mathrm{\mathbf{x}}^{(v)}|y, \mathrm{\mathbf{z}})$ and $q_{\varphi}(y|\mathfrak{X})$ correspond to the encoder, decoder and classifier, respectively.
For brevity, we omit the explicit dependencies on $\mathrm{\mathbf{x}}^{(v)}$, $y$ and $\mathrm{\mathbf{z}}$ for the moment variables mentioned above hereafter. In principle, $\bm{\mu}_{\theta^{(v)}}$, $\bm{\sigma}^{2}_{\theta^{(v)}}$, $\bm{\pi}_{\varphi}$, $\bm{\mu}_{\phi^{(v)}}$ and $\bm{\Sigma}_{\phi^{(v)}}$  can be implemented by various DNN models, e.g., multi-layer perceptrons and convolutional neural networks.

\subsubsection{Objective function}
In the semi-supervised setting, there are two lower bounds for the \textit{labeled} and \textit{unlabeled} cases, respectively.
The variational lower bound on the marginal likelihood for a single \textit{labeled} data point is
\begin{align} \label{L_XY}
\nonumber \log p_{\theta}(\mathfrak{X}, y) & \ge \mathbb{E}_{q_{\phi}(\mathrm{\mathbf{z}}|\mathfrak{X}, y)}\bigg[\log \frac{ p_{\theta}(\mathfrak{X}, y, \mathrm{\mathbf{z}})} {q_{\phi}(\mathrm{\mathbf{z}}|\mathfrak{X}, y)}\bigg]  \\
\nonumber & \ge \mathbb{E}_{q_{\phi}(\mathrm{\mathbf{z}}|\mathfrak{X},y)}\bigg[\sum_{v=1}^{2} \log p_{\theta^{(v)}}(\mathrm{\mathbf{x}}^{(v)}|y, \mathrm{\mathbf{z}}) + \log p(y)\\
\nonumber &\hspace{0.4cm} + \log p(\mathrm{\mathbf{z}}) \bigg] - \sum_{v=1}^2 \lambda^{(v)} \cdot \log \bigg(\sum_{l=1}^2 \lambda^{(l)} \cdot \omega_{v, l}\bigg) \\
 & \equiv -\mathcal{L}(\mathfrak{X}, y),
\end{align}
where $\omega_{v, l} = \mathcal{N}\big( \bm{\mu}_{\phi^{(v)}} | \bm{\mu}_{\phi^{(l)}},\ \bm{\Sigma}_{\phi^{(v)}} + \bm{\Sigma}_{\phi^{(l)}} \big)$. It should be noted that, the Shannon entropy $\mathbb{E}_{q_{\phi}(\mathrm{\mathbf{z}}|\mathfrak{X},y)}[- \log q_{\phi}(\mathrm{\mathbf{z}}|\mathfrak{X},y)]$ is hard to compute analytically, and we have used the Jensen's inequality to derive a lower bound of it (see Supplementary Material Section A for details).
For \textit{unlabeled} data point, the variational lower bound on the
marginal likelihood can be given by:
\begin{align} \label{ELBO_SMVAE_unlabeled}
\log p_{\theta}(\mathfrak{X}) & \ge \mathbb{E}_{q_{\varphi, \phi}(y, \mathrm{\mathbf{z}}|\mathfrak{X})}\bigg[\log \frac{ p_{\theta}(\mathfrak{X}, y, \mathrm{\mathbf{z}})} {q_{\varphi, \phi}(y, \mathrm{\mathbf{z}}|\mathfrak{X})}\bigg]  \\
\nonumber & = \mathbb{E}_{q_{\varphi}(y|\mathfrak{X})}\big[-\mathcal{L}(\mathfrak{X},y)- \log q_{\varphi}(y|\mathfrak{X})\big]
\equiv -\mathcal{U}(\mathfrak{X}),
\end{align}
with $q_{\varphi, \phi}(y, \mathrm{\mathbf{z}}|\mathfrak{X}) = q_{\varphi}(y|\mathfrak{X}) q_{\phi}(\mathrm{\mathbf{z}}|\mathfrak{X}, y)$.

Therefore, the objective function for the entire dataset is:
\begin{align}\label{obj_SMVAE_part}
\mathcal{J}_{\mathrm{SMVAE}}  = \underbrace{ \sum_{(\mathfrak{X},y) \in S_{l}}\mathcal{L}(\mathfrak{X},y)}_\text{\textit{labeled}} +  \underbrace{ \sum_{\mathfrak{X} \in S_{u}}\mathcal{U}(\mathfrak{X})}_\text{\textit{unlabeled}},
\end{align}
where $S_l$ and $S_u$ denote \textit{labeled} and \textit{unlabeled} dataset, respectively. The classification accuracy can be improved by introducing
an explicit classification loss for labeled data, and the extended objective function is  now:
\begin{align} \label{obj_SMVAE}
\mathcal{F}_{\mathrm{SMVAE}}  = \mathcal{J}_{\mathrm{SMVAE}} + \alpha \cdot \sum_{(\mathfrak{X},y) \in S_{l}}\big[-\log q_{\varphi}(y|\mathfrak{X})\big] \,,
\end{align}
where $\alpha$ is a weight parameter between generative and discriminative learning. We set $\alpha = c \cdot \frac{(N_l + N_u)}{N_l}$, where $c$ is a scaling constant, and $N_l$ and $N_u$ are the numbers of labeled and unlabeled data points in one minibatch, respectively.
Note that, the classifier $q_{\varphi}(y|\mathfrak{X})$ is also used at test phase for the prediction of unseen data.
Eq. (\ref{obj_SMVAE}) provides a unified objective function for optimizing the parameters of encoder, decoder and classifier networks.

\subsubsection{Parameter optimization}
Parameter optimization can be done jointly by using the stochastic backpropagation technique ~\cite{VAE,rezende2014stochastic}.
The reparameterization trick ~\cite{VAE,kingma2014semi} is a vital component, because it allows us to take derivative of  $\mathbb{E}_{q_{\phi}(\mathrm{\mathbf{z}}|\mathfrak{X},y)}$$[\log p_{\theta^{(v)}}(\mathrm{\mathbf{x}}^{(v)}|y, \mathrm{\mathbf{z}})]$ w.r.t. the variational parameters $\phi$.
However, the use of Gaussian mixture for variational posterior distribution $q_{\phi}(\mathrm{\mathbf{z}}|\mathfrak{X},y)$ makes it infeasible to apply the reparameterization trick directly.
It can be shown that, for any $v \in\{1, 2\}$, $\mathbb{E}_{q_{\phi}(\mathrm{\mathbf{z}}|\mathfrak{X},y)}[\log p_{\theta^{(v)}}(\mathrm{\mathbf{x}}^{(v)}|y, \mathrm{\mathbf{z}})]$  can be rewritten, using the location-scale transformation for the Gaussian distribution, as:
\begin{align}
\label{reparameterization}
& \mathbb{E}_{q_{\phi}(\mathrm{\mathbf{z}}|\mathfrak{X},y)}[\log p_{\theta^{(v)}}(\mathrm{\mathbf{x}}^{(v)}|y, \mathrm{\mathbf{z}})] \\
\nonumber  &  = \sum_{l=1}^2 \lambda^{(l)} \mathbb{E}_{\mathcal{N}(\bm{\epsilon}^{(l)}|\mathrm{\mathbf{0}},\mathrm{\mathbf{I}})}\bigg[\log p_{\theta^{(v)}} (\mathrm{\mathbf{x}}^{(v)}|y, \bm{\mu}_{\phi^{(l)}} + \mathrm{\mathbf{R}}_{\phi^{(l)}} \bm{\epsilon}^{(l)} ) \bigg],
\end{align}
where $\mathrm{\mathbf{R}}_{\phi^{(l)}}\mathrm{\mathbf{R}}_{\phi^{(l)}}^\top = \bm{\Sigma}_{\phi^{(l)}}$ and  $l\in\{1, 2\}$.
While the expectations on the right hand side still cannot be solved analytically, their gradients w.r.t. $\theta^{(v)}$, $\phi^{(l)}$ and $\lambda^{(l)}$ can be efficiently estimated using Monte-Carlo method (see Supplementary Material Section B for details).
The gradients of the objective function (Eq. (\ref{obj_SMVAE})) can then be computed by using the chain rule and the derived Monte-Carlo estimators.
\subsection{Handling Incomplete Data}
In the above discussion it is assumed that all modalities are available for every data point. In practice, however, many samples generally have incomplete modalities (i.e., with one or more missing modalities) ~\cite{wagner2011exploring}. In light of this, we further develop a semi-supervised incomplete multi-view classification algorithm (SiMVAE).
For simplicity, we assume only one view (either $\mathrm{\mathbf{x}}^{(1)}$ or $\mathrm{\mathbf{x}}^{(2)}$) is incomplete, though our model can be easily extended to more sophisticated cases.  We partition each data point into an observed view $\mathrm{\mathbf{x}}^{o}$ and a missing view $\mathrm{\mathbf{x}}^{m}$ (i.e., $\mathfrak{X} = \{\mathrm{\mathbf{x}}^{o}, \mathrm{\mathbf{x}}^{m}\}$).

\subsubsection{Generative model $P_3$}
In this setting, only a subset of the samples have complete views and corresponding labels. We regard both the unknown label $y$ and the missing view $\mathrm{\mathbf{x}}^{m}$ as latent variables.
Then our generative model $P_3$ (cf. Figure \ref{fig:MVAE}e) is defined as $p(y)p(\mathrm{\mathbf{z}}) p_{\theta^{m}}(\mathrm{\mathbf{x}}^{m}|y, \mathrm{\mathbf{z}}) p_{\theta^{o}}(\mathrm{\mathbf{x}}^{o}|y, \mathrm{\mathbf{z}}, \mathrm{\mathbf{x}}^{m}) $:
\begin{align}
p_{\theta^{m}}(\mathrm{\mathbf{x}}^{m}|y, \mathrm{\mathbf{z}}) & = \mathcal{N}\left(\bm{\mu}_{\theta^{m}}(y, \mathbf{z}),\ \mathrm{diag}(\bm{\sigma}^{2}_{\theta^{m}}(y, \mathbf{z})) \right),\\
\nonumber p_{\theta^{o}}(\mathrm{\mathbf{x}}^{o}|y, \mathrm{\mathbf{z}}, \mathrm{\mathbf{x}}^{m}) & = \mathcal{N}\left(\bm{\mu}_{\theta^{o}}(y, \mathbf{z}, \mathrm{\mathbf{x}}^{m}),\ \mathrm{diag}(\bm{\sigma}^{2}_{\theta^{o}}(y, \mathbf{z}, \mathrm{\mathbf{x}}^{m})) \right), \ \
\end{align}
where $p_{\theta^{m}}(\cdot)$ and $p_{\theta^{o}}(\cdot)$ are DNNs with parameters $\theta^{m}$ and $\theta^{o}$, respectively. $p(y)$ and $p(\mathrm{\mathbf{z}})$ are defined as in Eq. (\ref{SMVAE_generative}).

\subsubsection{Inference model $Q_3$}
As multi-modality emotional data are collected from the same subject, there must be some underlying relationships between modalities, though they focus on different information. Given the observed modality, the estimation of missing modality is feasible if we capture the relationships between modalities. Therefore, the inference model $Q_3$ (cf. Figure \ref{fig:MVAE}f) is defined as $q_{\psi}(\mathrm{\mathbf{x}}^{m}|\mathrm{\mathbf{x}}^{o})q_{\varphi}(y|\mathfrak{X})q_\phi(\mathrm{\mathbf{z}}|\mathfrak{X}, y)$, with
\begin{align}\label{conditional_density}
q_{\psi}(\mathrm{\mathbf{x}}^{m}|\mathrm{\mathbf{x}}^{o}) = \mathcal{N}\left(\bm{\mu}_{\psi}(\mathrm{\mathbf{x}}^{o}),\ \mathrm{diag}(\bm{\sigma}^{2}_{\psi}(\mathrm{\mathbf{x}}^{o})) \right),
\end{align}
where $q_{\psi}(\cdot)$ is a DNN with parameter $\psi$. $q_{\varphi}(y|\mathfrak{X})$ and $q_\phi(\mathrm{\mathbf{z}}|\mathfrak{X}, y)$ are defined as in Eq. (\ref{SMVAE_inference}). Intuitively, we formulate the missing view imputation as a conditional distribution estimation task (conditioned on the observed view). Compared with existing single imputation methods ~\cite{Chandar2016Correlational,Shang2017VIGAN,Luan2017Missing}, our model essentially performs infinite imputations and hence takes the uncertainty of the missing data into account.  To obtain a single imputation of $\mathrm{\mathbf{x}}^{m}$ rather than the full conditional distribution one can evaluate $\mathrm{\mathbf{x}}^{m} = \mathbb{E}[ q_{\psi}(\mathrm{\mathbf{x}}^{m}| \mathrm{\mathbf{x}}^{o})]$.

\subsubsection{Objective function}
In semi-supervised incomplete multi-view setting, there are four lower bounds for the \textit{labeled}-\textit{complete}, \textit{labeled}-\textit{incomplete}, \textit{unlabeled}-\textit{complete} and \textit{unlabeled}-\textit{incomplete} cases, respectively.

Similar to Eq. (\ref{L_XY}), the variational lower bound on the marginal likelihood for a single \textit{labeled}-\textit{complete} data point is
\begin{align}
\nonumber \log p_{\theta}(\mathfrak{X}, y)
& \ge \mathbb{E}_{q_{\phi}(\mathrm{\mathbf{z}}|\mathfrak{X}, y)}[\log p_{\theta^o}(\mathrm{\mathbf{x}}^{o}| \mathrm{\mathbf{x}}^{m}, y, \mathrm{\mathbf{z}})  + \log p(y) \\
\nonumber & \hspace{-1.7cm} + \log p_{\theta^m}(\mathrm{\mathbf{x}}^{m}|y, \mathrm{\mathbf{z}}) + \log p(\mathrm{\mathbf{z}}) ] - \sum_{v=1}^2 \lambda^{(v)} \cdot \log \bigg(\sum_{l=1}^2 \lambda^{(l)} \cdot \omega_{v, l}\bigg) \\
& \hspace{-1.7cm} \equiv -\mathcal{LC}(\mathfrak{X}, y),
\end{align}
where $\omega_{v, l} = \mathcal{N}\big( \bm{\mu}_{\phi^{(v)}} | \bm{\mu}_{\phi^{(l)}},\ \bm{\Sigma}_{\phi^{(v)}} + \bm{\Sigma}_{\phi^{(l)}} \big)$.
In the \textit{labeled}-\textit{incomplete} context, the variational lower bound on the
marginal likelihood for a single data point can be given by:
\begin{align}
& \log p_{\theta}(\mathrm{\mathbf{x}}^{o}, y)
\ge  \int_{\mathrm{\mathbf{z}}} \int_{\mathrm{\mathbf{x}}^{m}}  \log p_{\theta}(\mathfrak{X}, y, \mathrm{\mathbf{z}})\ d \mathrm{\mathbf{z}}\ d \mathrm{\mathbf{x}}^{m}\\
\nonumber & = \mathbb{E}_{q_{\psi}(\mathrm{\mathbf{x}}^{m}| \mathrm{\mathbf{x}}^{o})}\big[   -\mathcal{LC}(\mathfrak{X}, y)  - \log q_{\psi}(\mathrm{\mathbf{x}}^{m}| \mathrm{\mathbf{x}}^{o}) \big]
 \equiv -\mathcal{LI}(\mathrm{\mathbf{x}}^{o}, y).
\end{align}
The solution to $\mathbb{E}_{q_{\psi}(\mathrm{\mathbf{x}}^{m}| \mathrm{\mathbf{x}}^{o})}[ - \log q_{\psi}(\mathrm{\mathbf{x}}^{m}| \mathrm{\mathbf{x}}^{o})]$ is analytical since the conditional distribution $q_{\psi}(\mathrm{\mathbf{x}}^{m}| \mathrm{\mathbf{x}}^{o})$ is assumed to be a Gaussian (cf. Eq. (\ref{conditional_density})).
For \textit{unlabeled}-\textit{complete} data point, the variational lower bound on the
marginal likelihood can be obtained by
\begin{align} \label{unlabeled_complete}
\nonumber & \log p_{\theta}(\mathfrak{X})
\ge  \int_{\mathrm{\mathbf{z}}} \int_{y}  \log p_{\theta}(\mathfrak{X}, y, \mathrm{\mathbf{z}})\ d \mathrm{\mathbf{z}}\ d y\\
 & = \mathbb{E}_{q_{\varphi}(y|\mathfrak{X})}\big[   -\mathcal{LC}(\mathfrak{X}, y)  - \log q_{\varphi}(y|\mathfrak{X}) \big]
 \equiv -\mathcal{UC}(\mathfrak{X}).
\end{align}
For \textit{unlabeled}-\textit{incomplete} case, the variational lower bound on the
marginal likelihood can be given by:
\begin{align}
\nonumber & \log p_{\theta}(\mathrm{\mathbf{x}}^{o})
\ge  \int_{\mathrm{\mathbf{z}}} \int_{y} \int_{\mathrm{\mathbf{x}}^{m}} \log p_{\theta}(\mathfrak{X}, y, \mathrm{\mathbf{z}})\ d \mathrm{\mathbf{z}}\ d y\ d \mathrm{\mathbf{x}}^{m}\\
\nonumber & = \mathbb{E}_{q_{\psi}(\mathrm{\mathbf{x}}^{m}| \mathrm{\mathbf{x}}^{o})}\big\{\mathbb{E}_{q_{\varphi}(y|\mathfrak{X})}\big[   -\mathcal{LC}(\mathfrak{X}, y) - \log q_{\varphi}(y|\mathfrak{X}) \big] \\
& \hspace{3.2cm} - q_{\psi}(\mathrm{\mathbf{x}}^{m}| \mathrm{\mathbf{x}}^{o})\big\}
\equiv -\mathcal{UI}(\mathrm{\mathbf{x}}^{o}).
\end{align}
Comparing to Eq. (\ref{unlabeled_complete}) we see that aside from the explicit conditional distribution for unknown label $y$ we have added a conditional distribution $q_{\psi}(\mathrm{\mathbf{x}}^{m}| \mathrm{\mathbf{x}}^{o})$ for missing view $\mathrm{\mathbf{x}}^{m}$.

The objective function for all available data points is now:
\begin{align}\label{obj_SiMVAE_part}
 \nonumber  \mathcal{J}_{\mathrm{SiMVAE}} =  & \underbrace{ \sum_{(\mathfrak{X},y) \in S_{lc}}\mathcal{LC}(\mathfrak{X},y)}_\text{\textit{labeled}-\textit{complete}} + \underbrace{ \sum_{(\mathrm{\mathbf{x}}^{o},y) \in S_{li}}\mathcal{LI}(\mathrm{\mathbf{x}}^{o},y)}_\text{\textit{labeled}-\textit{incomplete}} \\
 & + \underbrace{ \sum_{\mathfrak{X} \in S_{uc}}\mathcal{UC}(\mathfrak{X})}_\text{\textit{unlabeled}-\textit{complete}} + \underbrace{ \sum_{\mathrm{\mathbf{x}}^{o} \in S_{ui}}\mathcal{UI}(\mathrm{\mathbf{x}}^{o})}_\text{\textit{unlabeled}-\textit{incomplete}}.
\end{align}
Model performance can be improved by introducing explicit imputation loss and classification loss for complete data and labeled data, respectively. Therefore, the final objective function is
\begin{align}\label{obj_SiMVAE}
 \nonumber \mathcal{F}_{\mathrm{SiMVAE}} & = \mathcal{J}_{\mathrm{SiMVAE}} + \alpha_1 \cdot \sum_{\mathfrak{X} \in S_{c}}\big[-\log q_{\psi}(\mathrm{\mathbf{x}}^{m}| \mathrm{\mathbf{x}}^{o})\big]\\
 & \hspace{1.67cm} + \alpha_2 \cdot \sum_{(\mathfrak{X},y) \in S_{l}}\big[-\log q_{\varphi}(y|\mathfrak{X})\big],
\end{align}
where $\alpha_1$ and $\alpha_2$ are weight parameters, $S_{c} = S_{lc} \cup S_{uc}$ and $S_{l} = S_{lc} \cup S_{li}$. We set $\alpha_1 = c_1 \cdot \frac{(N_c + N_i)}{N_c}$ and $\alpha_2 = c_2 \cdot \frac{(N_l + N_u)}{N_l}$, where $c_1$ and $c_2$ are scaling constants, and $N_c$, $N_i$, $N_l$ and $N_u$ are the numbers of complete, incomplete, labeled and unlabeled data in one minibatch, respectively.  Noted that the explicit
classification loss (i.e., last term in Eq. (\ref{obj_SiMVAE})) allows SiMVAE to use the partially observed category information to assist the generation of $\mathrm{\mathbf{x}}^{m}$ given $\mathrm{\mathbf{x}}^{o} $, which is more effective than the unsupervised imputation algorithms ~\cite{wang2015deep,Chandar2016Correlational}.
Similarly, Eq. (\ref{obj_SiMVAE}) can be optimized by using the stochastic backpropagation technique ~\cite{VAE,rezende2014stochastic}.

In principle, our SiMVAE can also handle multiple missing views simultaneously.  The formulas are omitted here since they can be derived straightforwardly by using multiple distinct conditional density functions $q_{\psi}(\mathrm{\mathbf{x}}^{m}| \mathrm{\mathbf{x}}^{o})$.

\subsubsection{Connections to auxiliary deep generative models}\label{SDGM+}
\citeauthor{maaloe2016auxiliary} ~\cite{maaloe2016auxiliary} proposed auxiliary DGMs (ADGM and SDGM) by defining the inference model as $q_{\psi}(\mathrm{\mathbf{a}}|\mathrm{\mathbf{x}}^{o})$$q_{\varphi}(y|\mathrm{\mathbf{a}}, \mathrm{\mathbf{x}}^{o})$$q_\phi(\mathrm{\mathbf{z}}|\mathrm{\mathbf{a}}, \mathrm{\mathbf{x}}^{o}, y)$, where $\mathrm{\mathbf{a}}$ is the  auxiliary variable introduced to make the variational distribution more expressive, and $q_{\psi}(\mathrm{\mathbf{a}}|\mathrm{\mathbf{x}}^{o})$$ = $$ \mathcal{N}(\bm{\mu}_{\psi}(\mathrm{\mathbf{x}}^{o}),$$\ \mathrm{diag}(\bm{\sigma}^{2}_{\psi}(\mathrm{\mathbf{x}}^{o})))$.
If $\mathrm{\mathbf{x}}^{m}$ is a totally unobservable variable in Figures \ref{fig:MVAE}e and \ref{fig:MVAE}f, similar to SDGM, SiMVAE becomes a two-layered stochastic model. Since the generative process is conditioned on the auxiliary variable, two-layered stochastic model is more flexible than ADGM ~\cite{maaloe2016auxiliary}. Standard ADGM and SDGM could not handle incomplete multi-view data. We endow them with this ability by forcing the inferred auxiliary variable $\mathrm{\mathbf{a}}$ close to $\mathrm{\mathbf{x}}^{m}$ on the set of complete data. E.g., we can obtain the objective function of SDGM$+$ by introducing an additional imputation loss to SDGM:
\begin{align}\label{obj_SDGM}
\mathcal{F}_{\mathrm{SDGM}+} & = \mathcal{J}_{\mathrm{SDGM}} + \alpha_{3} \cdot \sum_{\mathfrak{X} \in S_{c}}\big[-\log q_{\psi}(\mathrm{\mathbf{a}}| \mathrm{\mathbf{x}}^{o})\big],
\end{align}
where $\alpha_3$ is a regularization parameter, $\mathfrak{X}=\{\mathrm{\mathbf{x}}^{m}, \mathrm{\mathbf{x}}^{o}\}$ and $S_{c}$ denotes the set of complete data. $\mathcal{J}_{\mathrm{SDGM}}$ can be found in ~\cite{maaloe2016auxiliary}. Intuitively,  SDGM$+$ not only enjoys the advantages of SDGM (in terms of flexibility, convergence and performance), but also captures the relationships between views via the auxiliary inference model $q_{\psi}(\mathrm{\mathbf{a}}| \mathrm{\mathbf{x}}^{o})$.  However, SDGM$+$ sets a single Gaussian in the variational distribution $q_\phi(\mathrm{\mathbf{z}}|\mathrm{\mathbf{a}}, \mathrm{\mathbf{x}}^{o}, y)$, which may restrict its ability in multi-modality fusion.

\section{Experiments}
We conduct experiments on two multi-modal emotion datasets to demonstrate the effectiveness of the proposed framework.
\subsection{Datasets}
\noindent \textbf{SEED:}\ \ The SEED dataset ~\cite{Zheng2015Investigating} contains Electroencephalogram (EEG) and eye movement (Eye) signals from 9 subjects during watching 15 movie clips, where each movie clip lasts about 4 minutes long. The EEG signals were recorded from 62 channels and the Eye signals contained information about blink, saccade fixation and so on. We used the EEG and Eye data from 9 subjects across 3 sessions, totally 27 data files. For each data file, data from watching the 1-9 movie clips were used as training set, while data from watching the 10-12 movie clips were used as validation set and the rest (13-15) were used as testing set.

\smallskip
\noindent \textbf{DEAP:}The DEAP dataset  ~\cite{koelstra2012deap} contains EEG and peripheral physiological signals (PPS) from 32 subjects during watching 40 one-minute duration music videos. The EEG signals were recorded from 32 channels, whereas the PPS was recorded from 8 channels. The participants, using values from 1 to 9, rated each music video in terms of the levels of valence, arousal and so on. In our experiment, the valence-arousal space was divided into four quadrants according to the ratings. The threshold we used was 5, leading to four classes of data. Considering the variations of participants' ratings possibly associated with individual difference in rating scale, we discarded the samples whose ratings of arousal and valence are between 4 and 6. The dataset was randomly divided into 10-folds, where 8 folds for training, one fold for validation and the last fold for testing.

For SEED, we used the extracted differential entropy (DE) features and eye movement features (blink, saccade fixation and so on) ~\cite{lu2015combining}. For DEAP, following ~\cite{lu2015combining}, we split the time series data into many one-second non-overlapping segments, where each segment is treated as an instance. Then we extracted the DE features from EEG and PPS data instances. The DE features can be calculated in four frequency bands: theta (4-8Hz), alpha (8-14Hz), beta (14-31Hz), and gamma (31-45Hz), and we used all band's features. The details of the data used in our experiments were summarized in Table \ref{benchmark data sets}.
\begin{table}[!htbp]
\begin{center}
\scriptsize
\setlength\tabcolsep{2.0pt}
\setlength{\abovecaptionskip}{-3pt}
\vskip -0.001in
\begin{tabular}{|c|c|c|c|c|c|c|}
\hline
 dataset            & \#sample           &\#modality (\#dim.)        &\#training      & \#validation & \#test   & \#class     \\ \hline\hline
 SEED            & 22734                 &EEG(310), Eye(33)           & 13473          & 4725         & 4536   & 3       \\
 DEAP            & 21042                 &EEG(128), PPS(32)          & 16834          & 2104         & 2104    & 4       \\ \hline
\end{tabular}
\end{center}
\vskip -0.03in
\caption{Properties of the data used in experiments.}
\label{benchmark data sets}
\end{table}
\subsection{Semi-supervised Classification with Multi-Modality Emotional Data}
\subsubsection{Experimental setting}
To simulate SSL scenario, on both datasets, we randomly labeled different proportions of samples in the training set, and remained the rest samples in the training set unlabeled.  For transductive SSL, we trained models on the dataset consisting of the testing data and labeled data belonging to training set. For inductive SSL, we trained models on the entire training set consisting of the labeled and unlabeled data. For supervised learning, we trained models on the labeled data belonging to training set, and test their performance on the testing set.
We compared our SMVAE with a broad range of solutions, including MAE ~\cite{ngiam2011multimodal}, DCCA ~\cite{andrew2013deep}, DCCAE ~\cite{wang2015deep}, AMMSS ~\cite{Cai2013Heterogeneous}, AMGL ~\cite{nie2016parameter}, M2 ~\cite{kingma2014semi} and SDGM ~\cite{maaloe2016auxiliary}. For SMVAE, we considered multi-layer perceptrons as the type of inference and generative networks. On both datasets, we set the hidden architectures of the inference and generative networks for each view as `100-50-30' and `30-50-100', respectively, where `30' is the dimension of the latent variables.
We used the Adam optimizer~\cite{kingma2014adam} with a learning rate $\eta = 3\times10^{-4}$ in training. The scaling constant $c$ was selected from \{0.1, 0.5, 1\}.
For MAE, DCCA and DCCAE, we considered the same setups (network structure, learning rate, etc.) as our SMVAE. Furthermore, we used support vector machines (SVM) and transductive SVM (TSVM) for supervised learning and transductive SSL, respectively.
For AMGL, M2 and SDGM we used their default settings, and we evaluated M2's performance on each modality and the concatenation of all modalities, respectively.

\begin{table}[!ht]
\begin{center}
\scriptsize
\setlength\tabcolsep{3.0pt}
\renewcommand{\multirowsetup}{\centering}
\vskip -0.1in
\begin{tabular}{|l|l|c|c|c|}
  \hline
  SEED data &Algorithms & 1\% labeled & 2\% labeled & 3\% labeled\\
  \hline
  \hline
  \multirow{3}{1.5cm}{Supervised learning}
  &MAE+SVM ~\cite{ngiam2011multimodal}                      & .814$\pm$.031            & .896$\pm$.024           & .925$\pm$.024 \\
  &DCCA+SVM ~\cite{andrew2013deep}                      & .809$\pm$.035            & .891$\pm$.035           & .923$\pm$.028 \\
  &DCCAE+SVM ~\cite{wang2015deep}                      & .819$\pm$.036            & .893$\pm$.034           & .923$\pm$.027 \\
  \hline
  \multirow{6}{1.5cm}{Transductive SSL}
  &AMMSS ~\cite{Cai2013Heterogeneous}                       & .731$\pm$.055            & .839$\pm$.036           & .912$\pm$.018  \\
  &AMGL  ~\cite{nie2016parameter}                       & .711$\pm$.047            & .817$\pm$.023           & .886$\pm$.028 \\
  &MAE+TSVM ~\cite{ngiam2011multimodal}                    & .818$\pm$.035            & .910$\pm$.025           & .931$\pm$.026 \\
  &DCCA+TSVM ~\cite{andrew2013deep}                    & .811$\pm$.031            & .903$\pm$.024           & .928$\pm$.021 \\
  &DCCAE+TSVM ~\cite{wang2015deep}                  & .823$\pm$.040            & .907$\pm$.027           & .929$\pm$.023 \\
  &SMVAE                    & \textbf{.861}$\pm$.037   & \textbf{.931}$\pm$.020  & \textbf{.960}$\pm$.021 \\
  \hline
  \multirow{4}{1.5cm}{Inductive SSL}
  &M2 (Eye) ~\cite{kingma2014semi}                  & .753$\pm$.024            & .849$\pm$.055          & .899$\pm$.049 \\
  &M2 (EEG) ~\cite{kingma2014semi}                  & .768$\pm$.041            & .861$\pm$.040          & .919$\pm$.026 \\
  &M2 (Concat.) ~\cite{kingma2014semi}              & .803$\pm$.035            & .876$\pm$.043          & .926$\pm$.044  \\
  &SDGM (Concat.) ~\cite{maaloe2016auxiliary}       & .819$\pm$.034            & .893$\pm$.042          & .932$\pm$.041  \\
  &SMVAE                                            & \textbf{.880}$\pm$.033   & \textbf{.955}$\pm$.020  & \textbf{.968}$\pm$.015 \\
  \hline
\end{tabular}
\vskip 0.07in
\begin{tabular}{|l|l|c|c|c|}
  \hline
  DEAP data &Algorithms & 1\% labeled & 2\% labeled & 3\% labeled\\
  \hline
  \hline
  \multirow{3}{1.5cm}{Supervised learning}
  &MAE+SVM ~\cite{ngiam2011multimodal}                      & .353$\pm$.027            & .387$\pm$.014           & .411$\pm$.016 \\
  &DCCA+SVM ~\cite{andrew2013deep}                     & .359$\pm$.016            & .400$\pm$.014           & .416$\pm$.018 \\
  &DCCAE+SVM ~\cite{wang2015deep}                    & .361$\pm$.023            & .403$\pm$.017           & .419$\pm$.013 \\
  \hline
  \multirow{6}{1.5cm}{Transductive SSL}
  &AMMSS ~\cite{Cai2013Heterogeneous}                       & .303$\pm$.029            & .353$\pm$.024           & .386$\pm$.014  \\
  &AMGL  ~\cite{nie2016parameter}                       & .291$\pm$.027            & .341$\pm$.021           & .367$\pm$.019 \\
  &MAE+TSVM ~\cite{ngiam2011multimodal}                    & .376$\pm$.025            & .403$\pm$.031           & .417$\pm$.026 \\
  &DCCA+TSVM ~\cite{andrew2013deep}                    & .379$\pm$.021            & .408$\pm$.024           & .421$\pm$.017 \\
  &DCCAE+TSVM ~\cite{wang2015deep}                  & .384$\pm$.022            & .412$\pm$.027           & .425$\pm$.021 \\
  &SMVAE                   & \textbf{.424}$\pm$.020   & \textbf{.441}$\pm$.013  & \textbf{.456}$\pm$.013 \\
  \hline
  \multirow{4}{1.5cm}{Inductive SSL}
  &M2 (PPS) ~\cite{kingma2014semi}                & .366$\pm$.024            & .389$\pm$.048          & .402$\pm$.034 \\
  &M2 (EEG) ~\cite{kingma2014semi}                 & .374$\pm$.019            & .397$\pm$.013          & .407$\pm$.016 \\
  &M2 (Concat.) ~\cite{kingma2014semi}              & .383$\pm$.019            & .404$\pm$.016          & .416$\pm$.015  \\
  &SDGM (Concat.) ~\cite{maaloe2016auxiliary}       & .389$\pm$.019            & .411$\pm$.017          & .423$\pm$.015  \\
  &SMVAE                   & \textbf{.421}$\pm$.017   & \textbf{.439}$\pm$.015  & \textbf{.451}$\pm$.013 \\
  \hline
\end{tabular}
\end{center}
\caption{Comparison of classification accuracies with different proportions of labeled training samples. }
\label{accuracy}
\end{table}
\subsubsection{Classification accuracy with very few labels}
Table \ref{accuracy} presents the classification accuracies of all methods on SEED and DEAP datasets. The proportions of labeled samples in the training set vary from  $1\%$ to $3\%$. Results (mean$\pm$std) were averaged over 20 independent runs. Several observations can be drawn as follows.
First, the average accuracy of SMVAE significantly surpasses the baselines in all cases.
Second, by examining SMVAE against supervised learning approaches trained on very limited labeled data, we can find that SMVAE always outperforms them. This encouraging result shows that SMVAE can effectively  leverage the useful information from unlabeled data.
Third,  multi-view semi-supervised algorithms AMMSS and AMGL perform worst in all cases. We attribute this to the fact that graph-based shallow models AMMSS and AMGL cannot extract the deep features from the original data.
Fourth, the performances of three TSVM-based semi-supervised methods are moderate.
Finally, compared with the single-view methods M2 and SDGM, our multi-view method is more effective in integrating multiple modalities.

\subsubsection{Flexibility and stability}
\begin{figure}[!htbp]
\centering
  \subfigure[] {\includegraphics[height=0.8in,width=1.1in]{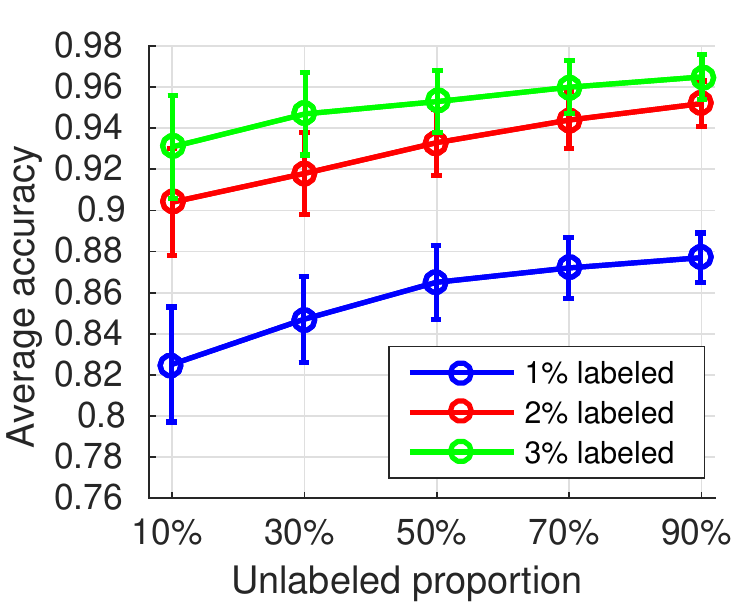}}
  \subfigure[] {\includegraphics[height=0.8in,width=1.05in]{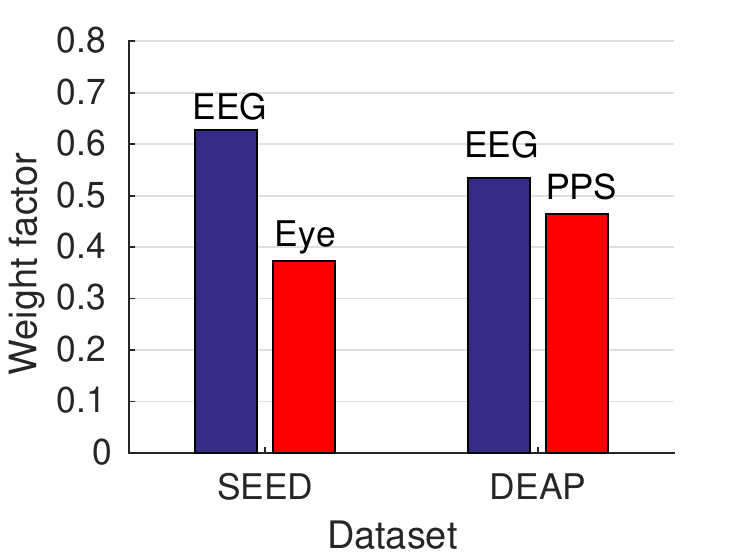}}
  \subfigure [] {\includegraphics[height=0.8in,width=1.1in]{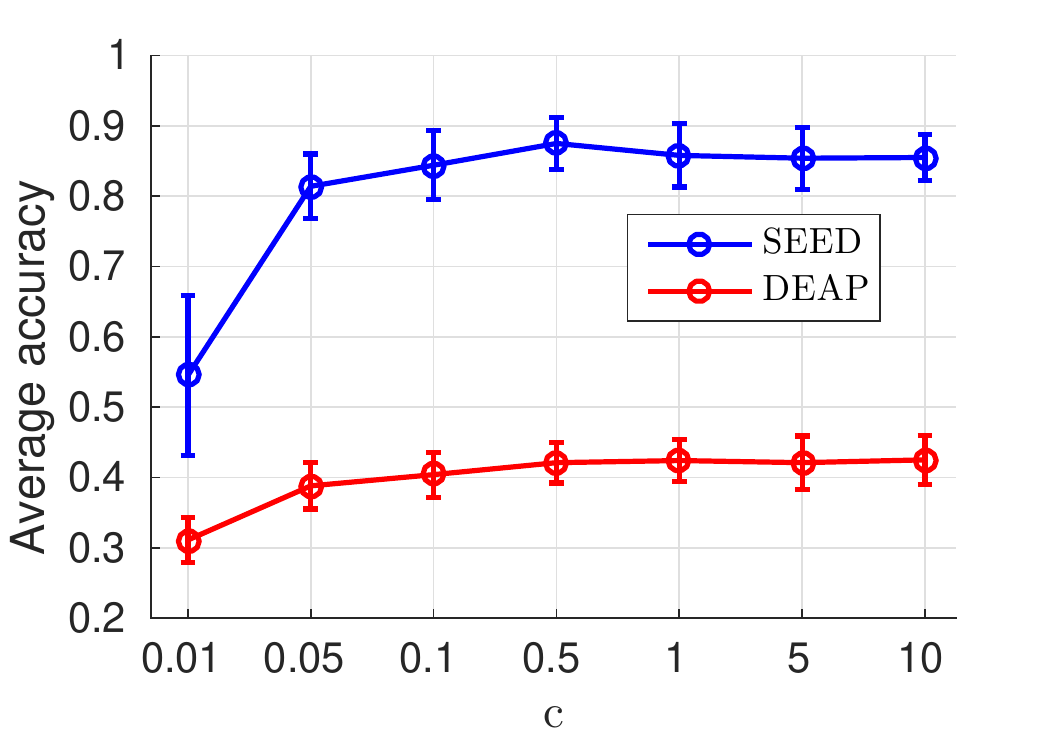}}
\caption{For Inductive SMVAE: (a) performance with different proportions of unlabeled training samples on SEED dataset,\ (b) learned weight factors,\ (c) the impact of scaling constant $c$.}
\label{fig:sensitive}
\end{figure}

The proportion of unlabeled samples in the training set will affect the performance of semi-supervised models. Figure \ref{fig:sensitive}a shows the changes of inductive SMVAE's average accuracy on SEED with different proportions of unlabeled samples in the training set. We can observe that the unlabeled samples can effectively boost the classification accuracy of SMVAE.
Instead of treating each modality equally, SMVAE can weight each modality and perform classification simultaneously. Figure \ref{fig:sensitive}b
shows the learned weight factors by inductive SMVAE on both datasets ($1\%$ labeled).  From it, we can observe that EEG modality has the highest weight on both datasets, which is consistent with single modality's performance of M2 shown in Table \ref{accuracy} and the results in previous work~\cite{lu2015combining}.
The scaling constant $c$ controls the weight of discriminative learning in SMVAE. Figure \ref{fig:sensitive}c shows the performance of inductive SMVAE with different $c$ values ($1\%$ labeled). From it, we can find that the scaling constant $c$ can be chosen from \{0.1, 0.5, 1\}, where SMVAE achieves good results.
\subsection{Semi-supervised Learning with Incomplete Multi-Modality Data}
\subsubsection{Experimental setting}
To simulate the incomplete data setting, we randomly selected a fraction of instances (from both labeled and unlabeled training data) to be unpaired examples, i.e., they are described by only one modality, and the remaining ones appear in both modalities. We varied the fraction of missing data from 10\% to 90\% with an interval of 20\%, while no missing data in validation and testing sets. In our experiment, we assumed the Eye modality of SEED and the PPS modality of DEAP are incomplete.

There are two main solutions for semi-supervised classification of incomplete multi-view data. One way is to complete the missing view firstly in an unsupervised way, and then conduct semi-supervised classification. Another way is to integrate missing view imputation and semi-supervised classification into an end-to-end learning framework. We compared our (inductive) SiMVAE algorithm with these two ways. Specifically, we compared SiMVAE with SoftImputeALS ~\cite{Hastie2015Matrix}, DCCAE ~\cite{wang2015deep}, CorrNet ~\cite{Chandar2016Correlational}, CoNet ~\cite{Quanz2012CoNet} and SDGM$+$ (a variant of SDGM ~\cite{maaloe2016auxiliary}, cf. Section \ref{SDGM+}).
For SoftImputeALS, DCCAE and CorrNet, we first estimated the missing modalities by using the authors' implementation, and then conducted semi-supervised classification by using our (inductive) SMVAE algorithm. For CoNet and SDGM$+$, we conducted missing modality imputation and semi-supervised classification simultaneously based on our own implementations.
Additionally, we also compared SiMVAE with the following two baselines:
1) SiMVAE with complete data (FullData, i.e., no missing modality for any training instances), which can be regarded as a upper bound of SiMVAE;
2) SiMVAE with only paired data (PartialData, i.e., we simply discard those incomplete samples in training process), which can be regarded as a lower bound of SiMVAE. These two bounds define the potential range of SiMVAE's performance.
For SiMVAE, both $c_1$ and $c_2$ were selected from \{0.1, 0.5, 1\}. For SDGM$+$, we selected the regularization parameter $\alpha_3$ from $\{1e-3, 1e-2, \cdots, 1e3\}$.
\begin{figure}[!htbp]
\centering
  \subfigure[SEED 1\% labeled] {\includegraphics[height=1.03in,width=1.1in]{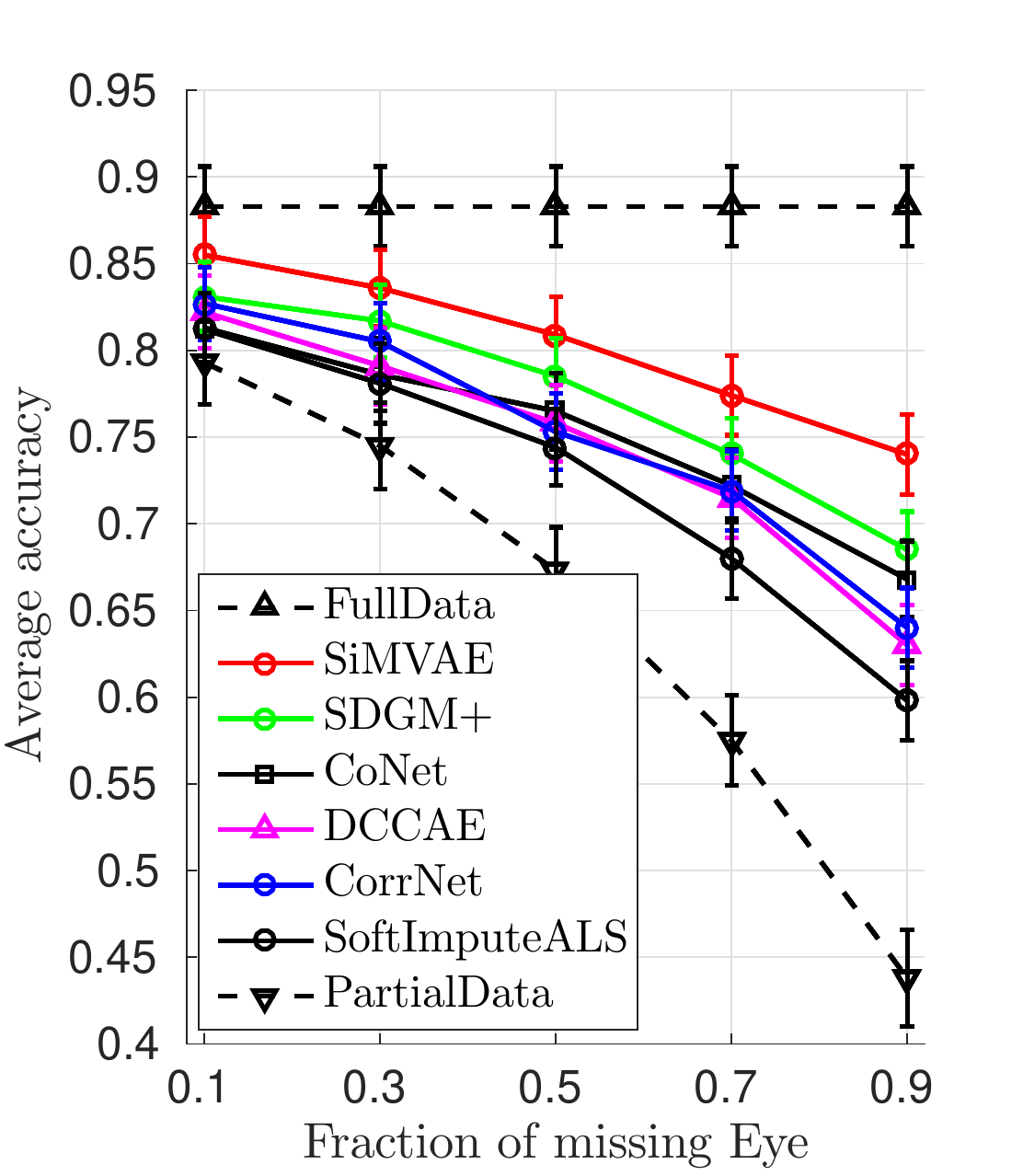}}
  \hspace {-0.25cm}
  \subfigure [SEED 2\% labeled] {\includegraphics[height=1.03in,width=1.1in]{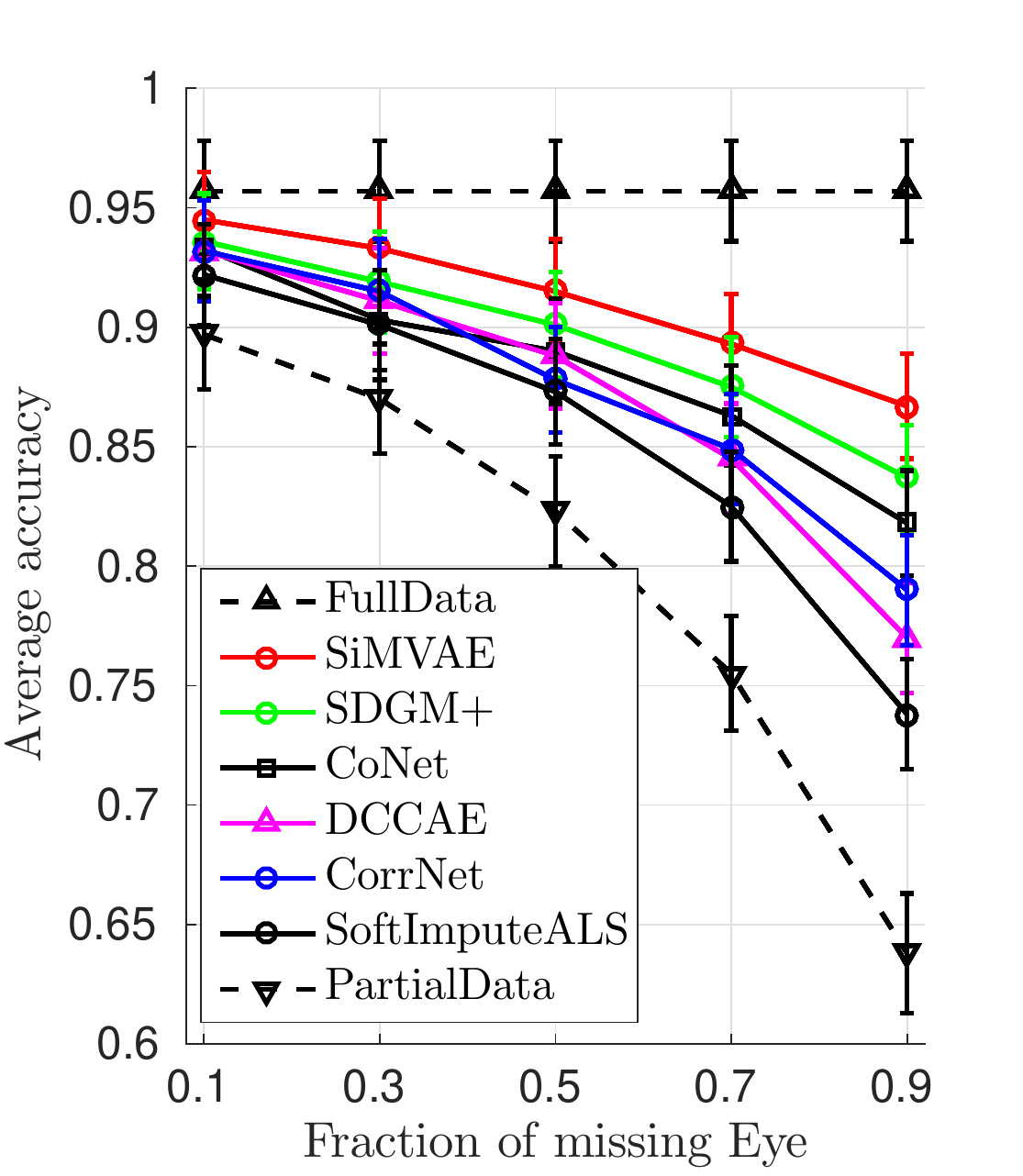}}
  \hspace {-0.25cm}
  \subfigure[SEED 3\% labeled] {\includegraphics[height=1.03in,width=1.1in]{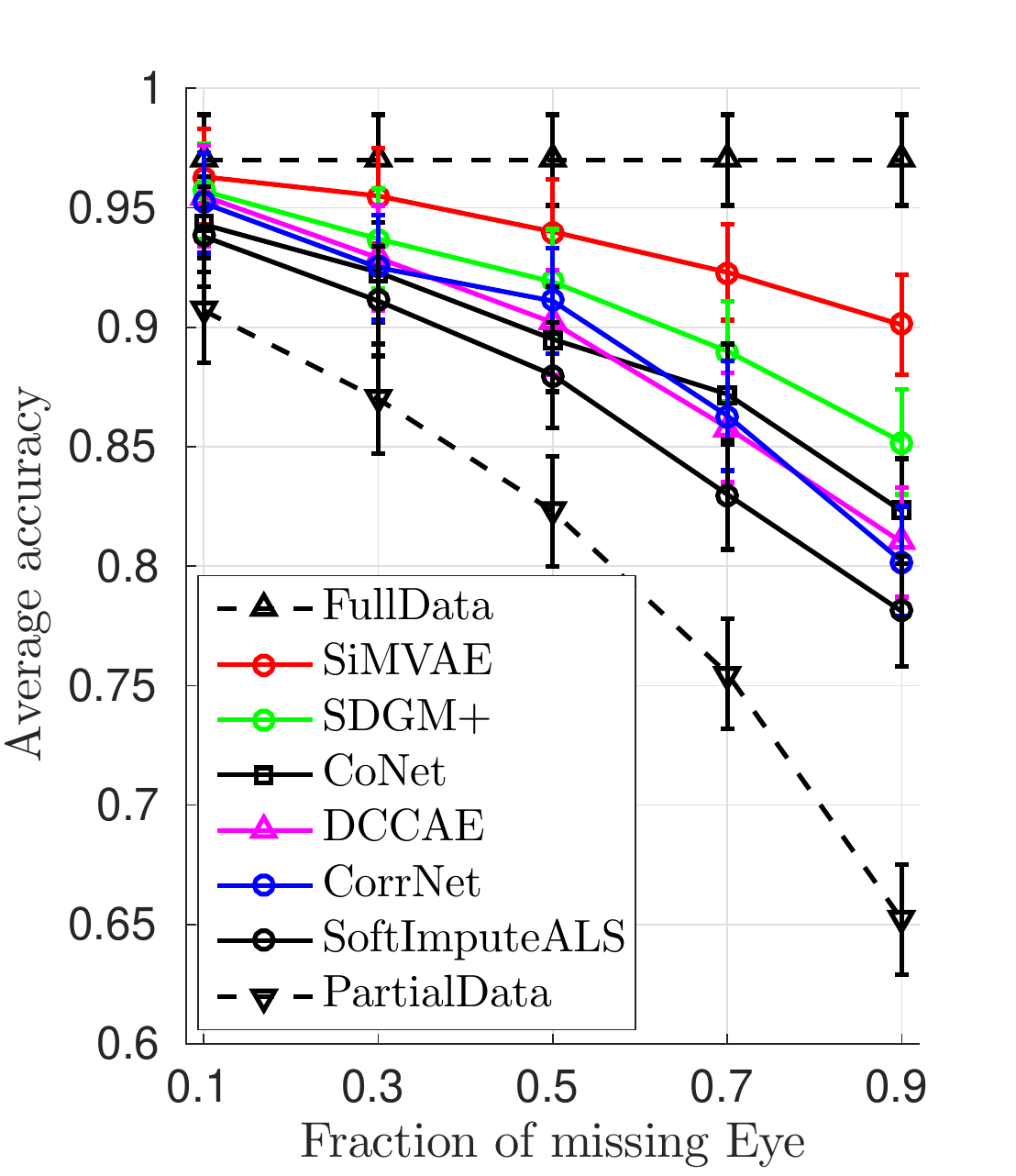}}
  \hspace {-0.25cm}
  \subfigure[DEAP 1\% labeled] {\includegraphics[height=1.03in,width=1.1in]{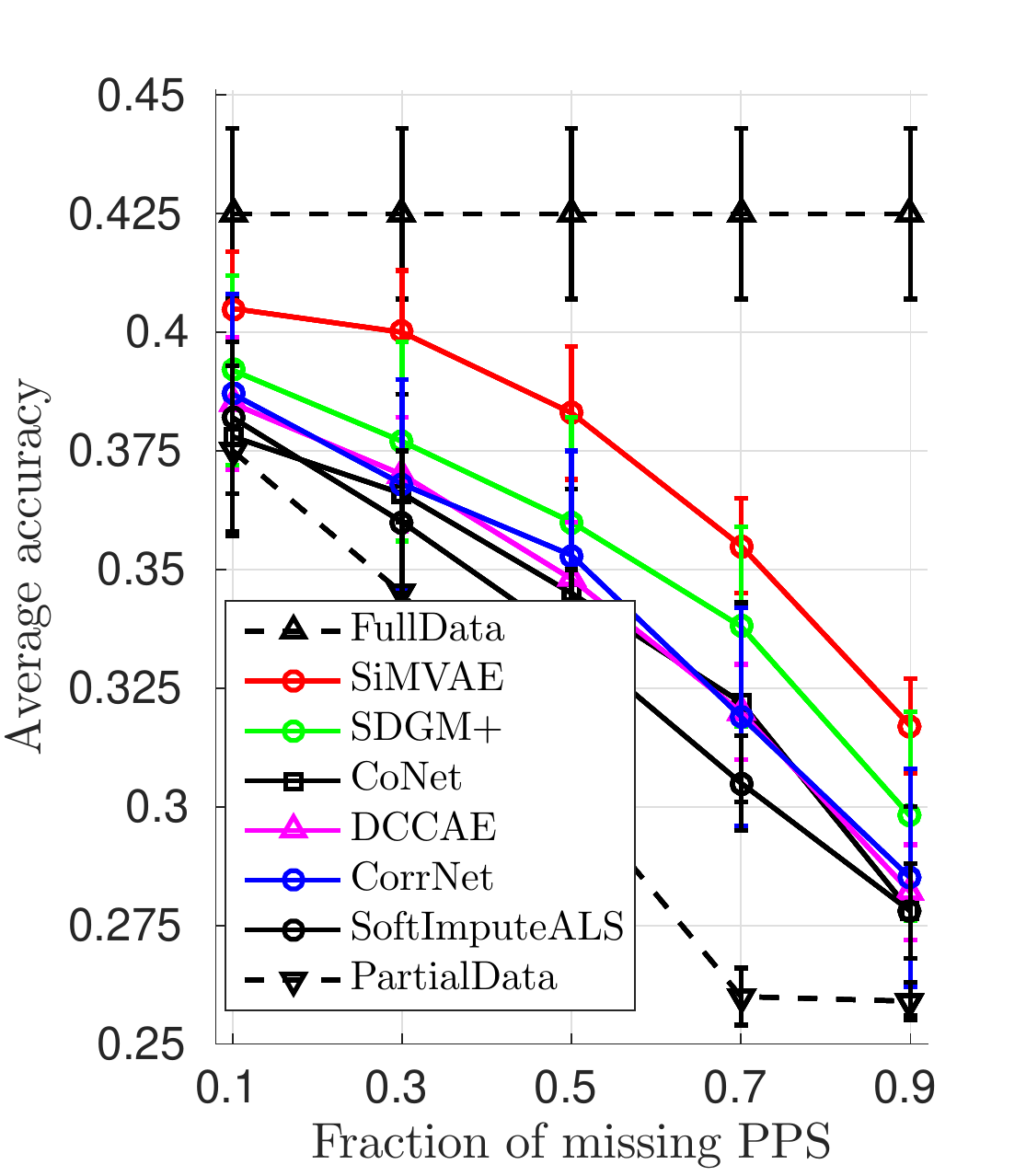}}
  \hspace {-0.25cm}
  \subfigure [DEAP 2\% labeled] {\includegraphics[height=1.03in,width=1.1in]{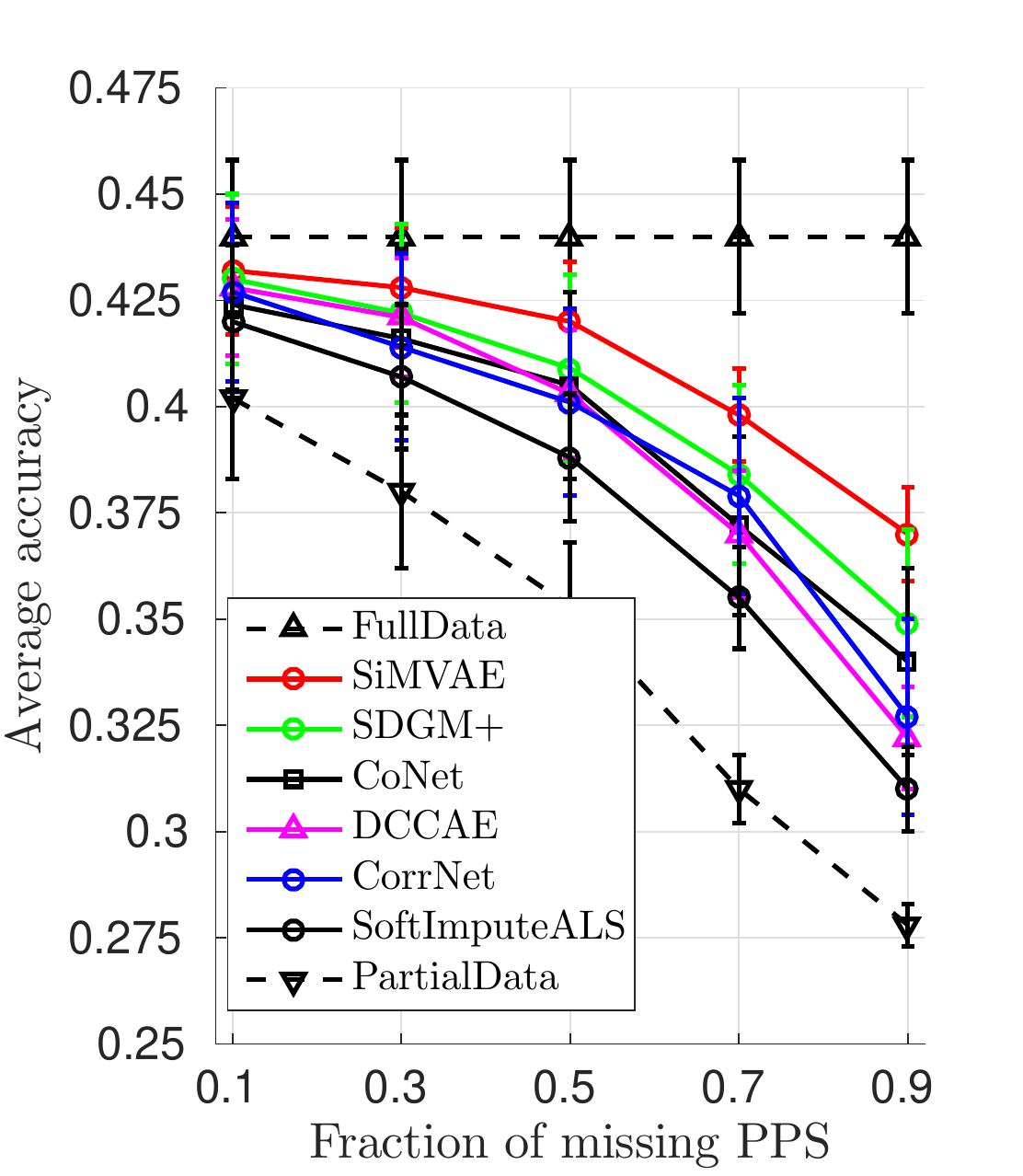}}
  \hspace {-0.25cm}
  \subfigure[DEAP 3\% labeled] {\includegraphics[height=1.03in,width=1.1in]{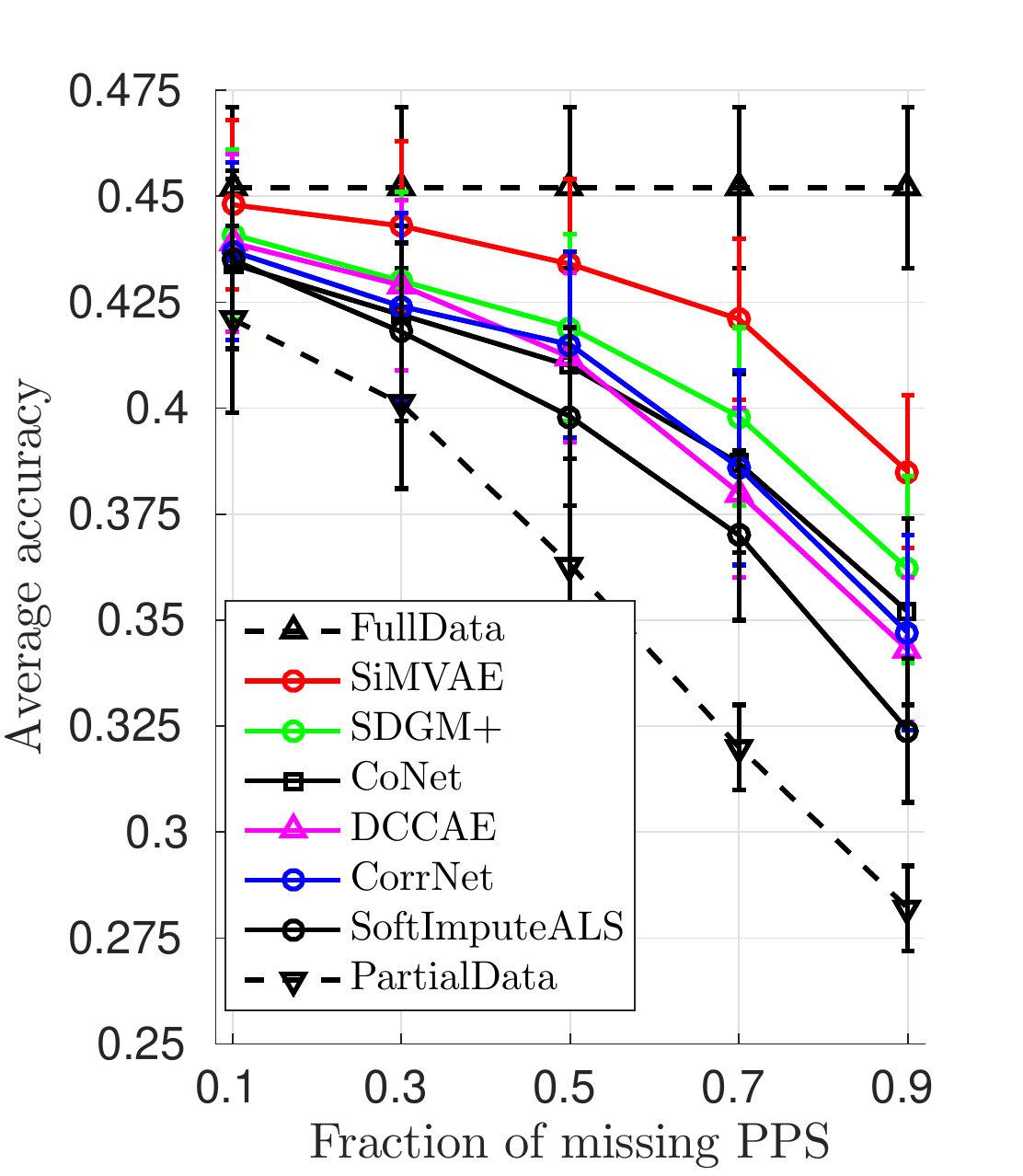}}
\caption{Comparison of recognition accuracies with different fractions of missing data and labeled data.}
\label{fig:seed_deap_unpaired_incre}
\end{figure}
\subsubsection{Semi-supervised classification}
The performance of our SiMVAE and the compared methods was shown in Figure \ref{fig:seed_deap_unpaired_incre}, where each point on every curve is an average over 20 independent trials. 
From Figure \ref{fig:seed_deap_unpaired_incre}, it is seen that SiMVAE consistently outperforms the compared methods. Compared with the two-stage methods (SoftImputeALS, DCCAE and CorrNet), the advantage of SiMVAE is significant, especially when there are sufficient labeled data (3\%). This is because SiMVAE can make good use of the available category information to generate more informative modalities, which in turn will improve classification performance. Whereas the two-stage methods couldn't obtain the global optimal results. Also, SiMVAE shows obvious advantage over the semi-supervised methods CoNet and SDGM$+$. This may be because CoNet and SDGM$+$ are not designed to integrate multiple modalities.
Moreover, SiMVAE has been successful even when a high percentage of samples are incomplete.
Specifically, SiMVAE with even about 50\% incomplete samples achieves comparable results to the fully complete case (FullData). With fractions lower than that, we observe that SiMVAE roughly reached FullData's performance, especially when the labeled data are sufficient.
Finally, SiMVAE's performance is more closer to FullData than to PartialData, which indicates the effectiveness of SiMVAE in learning from incomplete data.

\subsubsection{Missing modality imputation}
\begin{figure}[!htbp]
\centering
  \subfigure[SEED 1\% labeled] {\includegraphics[height=0.85in,width=1.13in]{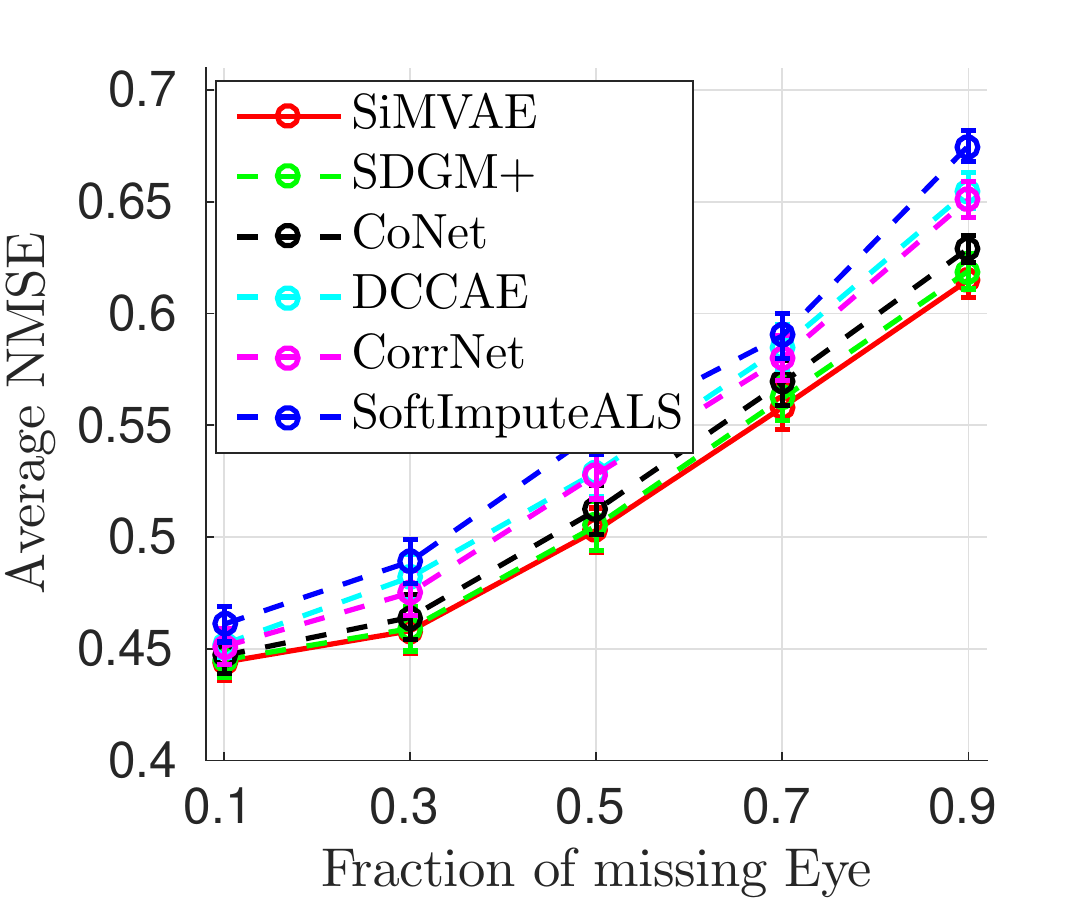}}
  \hspace {-0.27cm}
  \subfigure [SEED 2\% labeled] {\includegraphics[height=0.85in,width=1.13in]{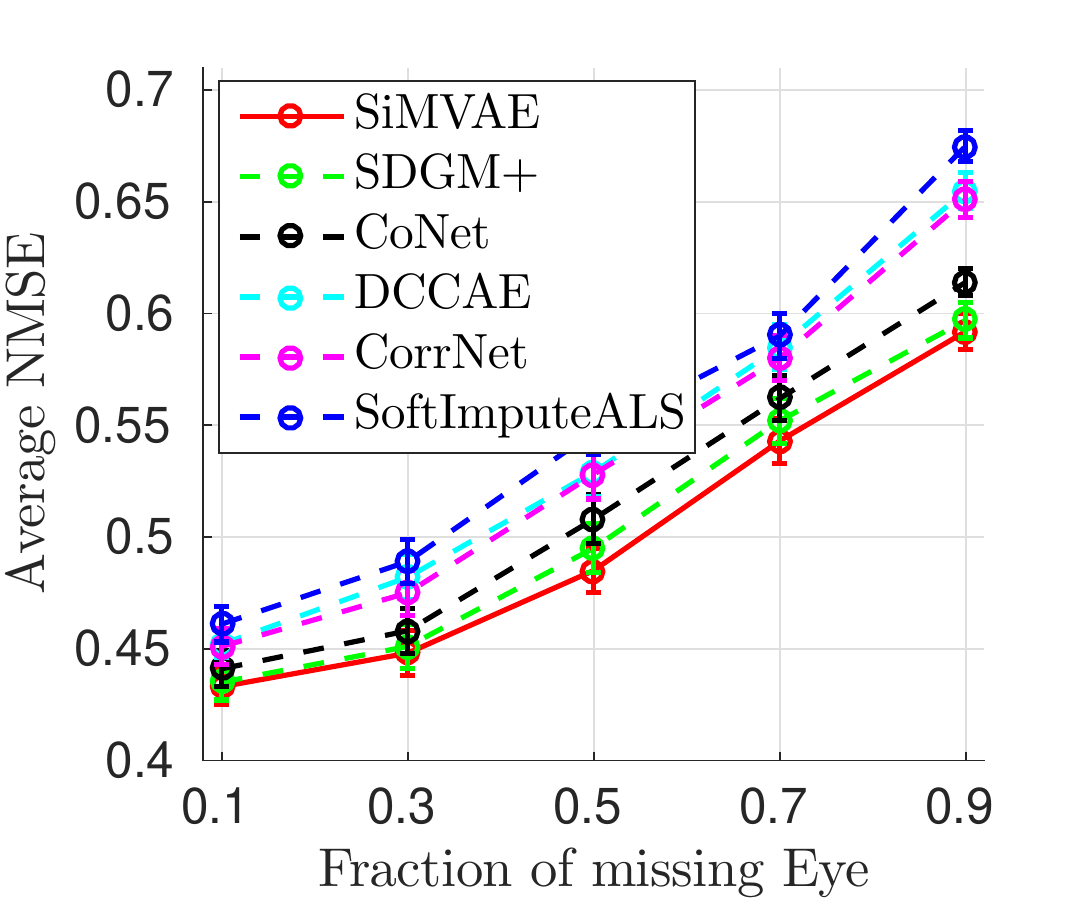}}
  \hspace {-0.27cm}
  \subfigure[SEED 3\% labeled] {\includegraphics[height=0.85in,width=1.13in]{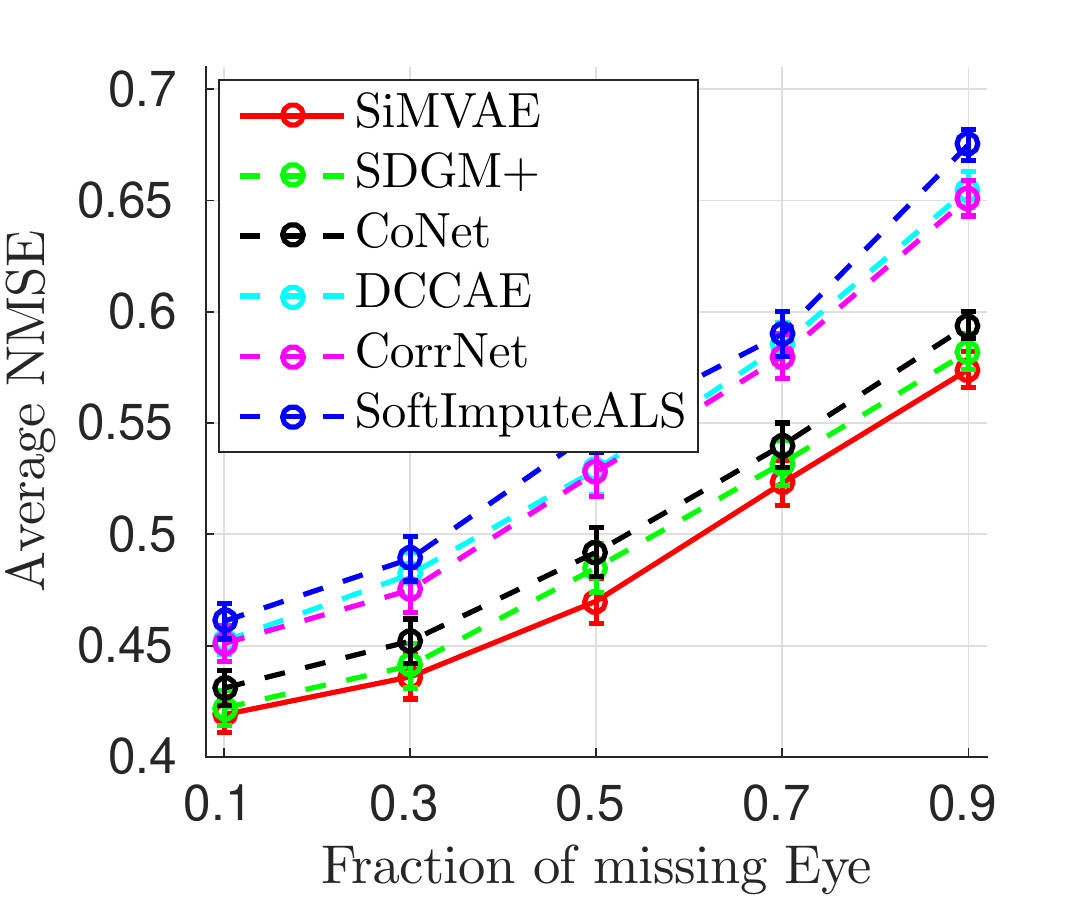}}
  \hspace {-0.27cm}
  \subfigure[DEAP 1\% labeled] {\includegraphics[height=0.85in,width=1.13in]{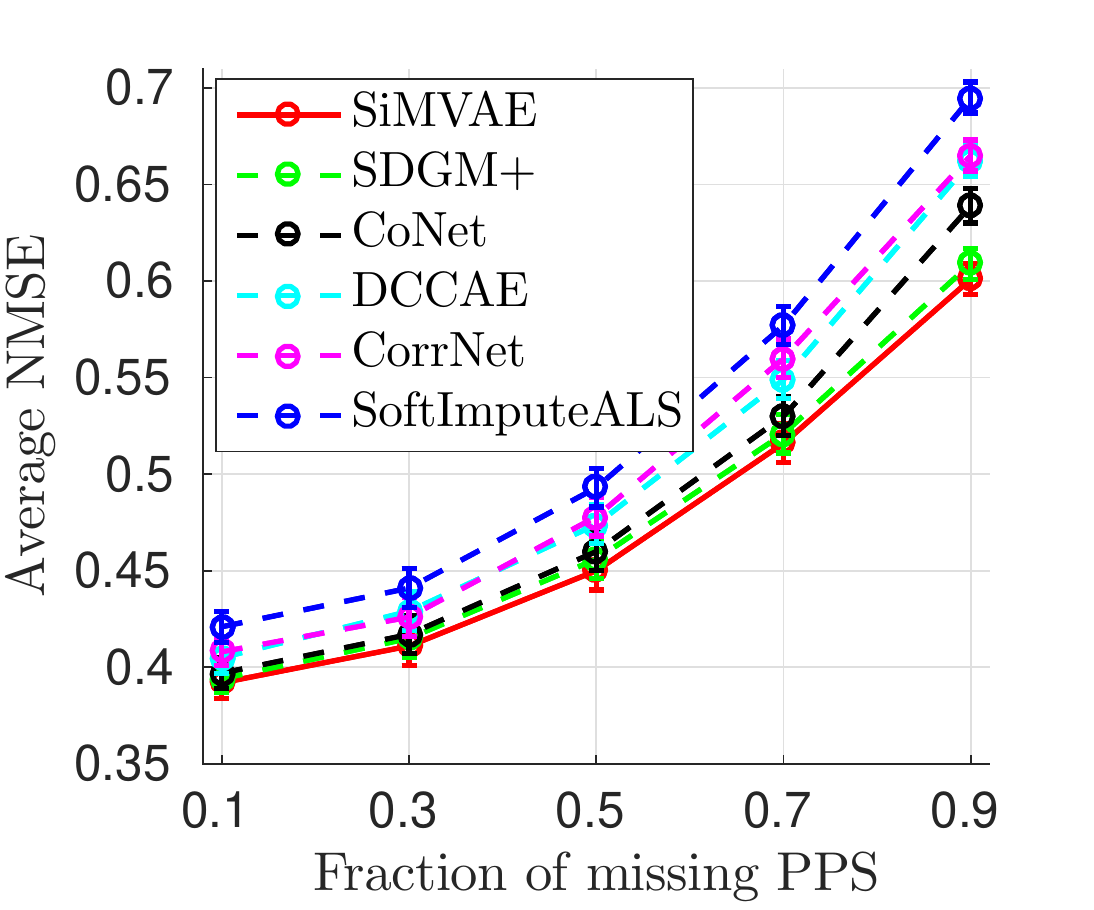}}
  \hspace {-0.27cm}
  \subfigure [DEAP 2\% labeled] {\includegraphics[height=0.85in,width=1.13in]{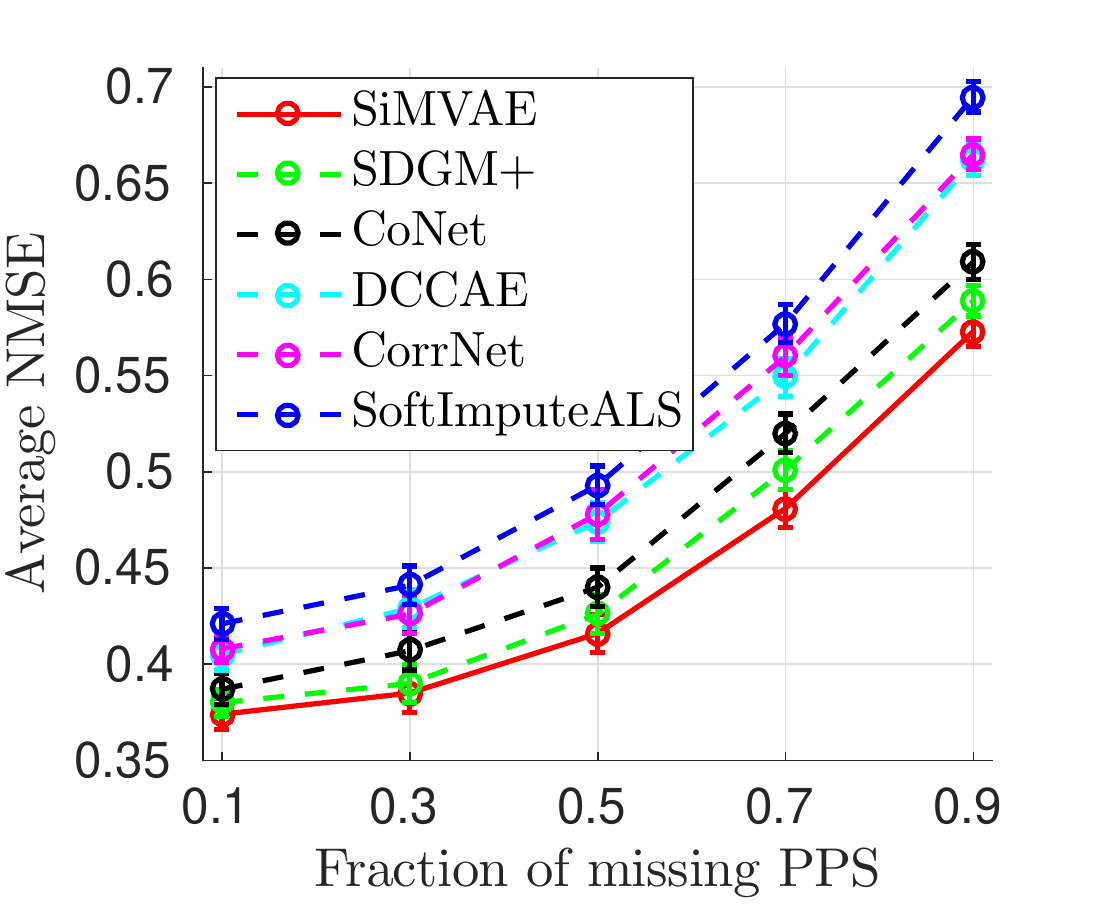}}
  \hspace {-0.27cm}
  \subfigure[DEAP 3\% labeled] {\includegraphics[height=0.85in,width=1.13in]{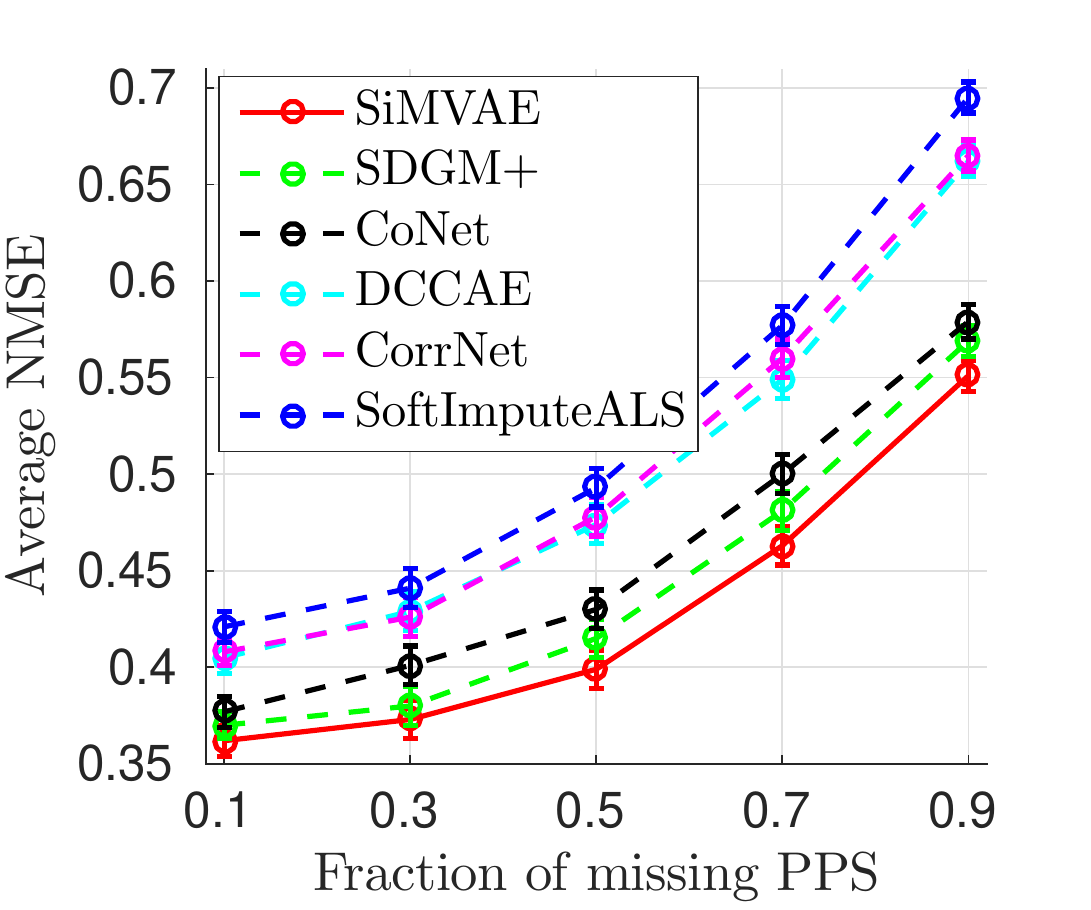}}
\vskip -0.18in
\caption{Comparison of imputation errors with different fractions of missing data and labeled data.}
\label{fig:imputation_mse}
\end{figure}

Since the quality of recovered missing modalities directly affects the classification results, we also evaluated the performance of missing modality imputation for all methods. For SiMVAE and SDGM$+$, we obtained the single imputation of $\mathrm{\mathbf{x}}^{m}$ by evaluating the conditional mean ($\mathrm{\mathbf{x}}^{m} = \mathbb{E}[ q_{\psi}(\mathrm{\mathbf{x}}^{m}| \mathrm{\mathbf{x}}^{o})]$). We used the Normalized Mean Squared Error (NMSE) to measure the relative distance between the original and the recovered modalities. $\mathrm{NMSE} = \frac{\|\mathrm{\mathbf{X}} - \hat{\mathrm{\mathbf{X}}}\|_F}{\|\mathrm{\mathbf{X}}\|_F}$, where $\mathrm{\mathbf{X}}$ and $\hat{\mathrm{\mathbf{X}}}$ are the original and the recovered data matrices, respectively. $\| \cdot \|_F$ demotes the Frobenious norm. Figure \ref{fig:imputation_mse} shows the experimental results.

\begin{figure}[!htbp]
\includegraphics[height=0.6in,width=3.3in]{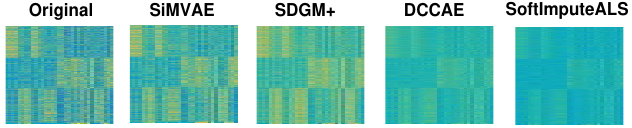}
\caption{Visualization of the original and the recovered data matrices on SEED dataset. Each row of each panel is an instance of the missing modality.}
\label{fig:imputation_vis}
\end{figure}

From Figure \ref{fig:imputation_mse}, it can be seen that as the fraction of missing data increases, the relative distance between the original modalities and the recovered modalities increases.  Further, the semi-supervised imputation methods (SiMVAE, CoNet and SDGM$+$) consistently outperforms the unsupervised imputation methods (SoftImputeALS, DCCAE and CorrNet), and increasing the number of labeled training data improves the imputation performance of semi-supervised methods. This demonstrates that the category information plays an important role in missing modality imputation. SoftImputeALS shows the worst performance, which verifies that matrix completion method is not suitable for missing modality imputation. CoNet and SDGM$+$ obtain comparable imputation errors to SiMVAE. This indicates that their moderate classification performance in Figure \ref{fig:seed_deap_unpaired_incre} may be caused by their inability in modality fusion. Except for SiMVAE and SDGM$+$, other methods ignore the uncertainty of the missing view, which also limits their imputation performance.
To compare the imputation performance more intuitively, we visualize the original and recovered data matrices in Figure \ref{fig:imputation_vis} (on SEED, 3\% labeled and 10\% missing Eye). From it, we see that SiMVAE recovered more individual characteristics of the original data matrix than other methods.

\subsubsection{Sensitivity analysis}
\begin{figure}[!htbp]
\centering
  \subfigure[SEED dataset] {\includegraphics[height=0.85in,width=1.53in]{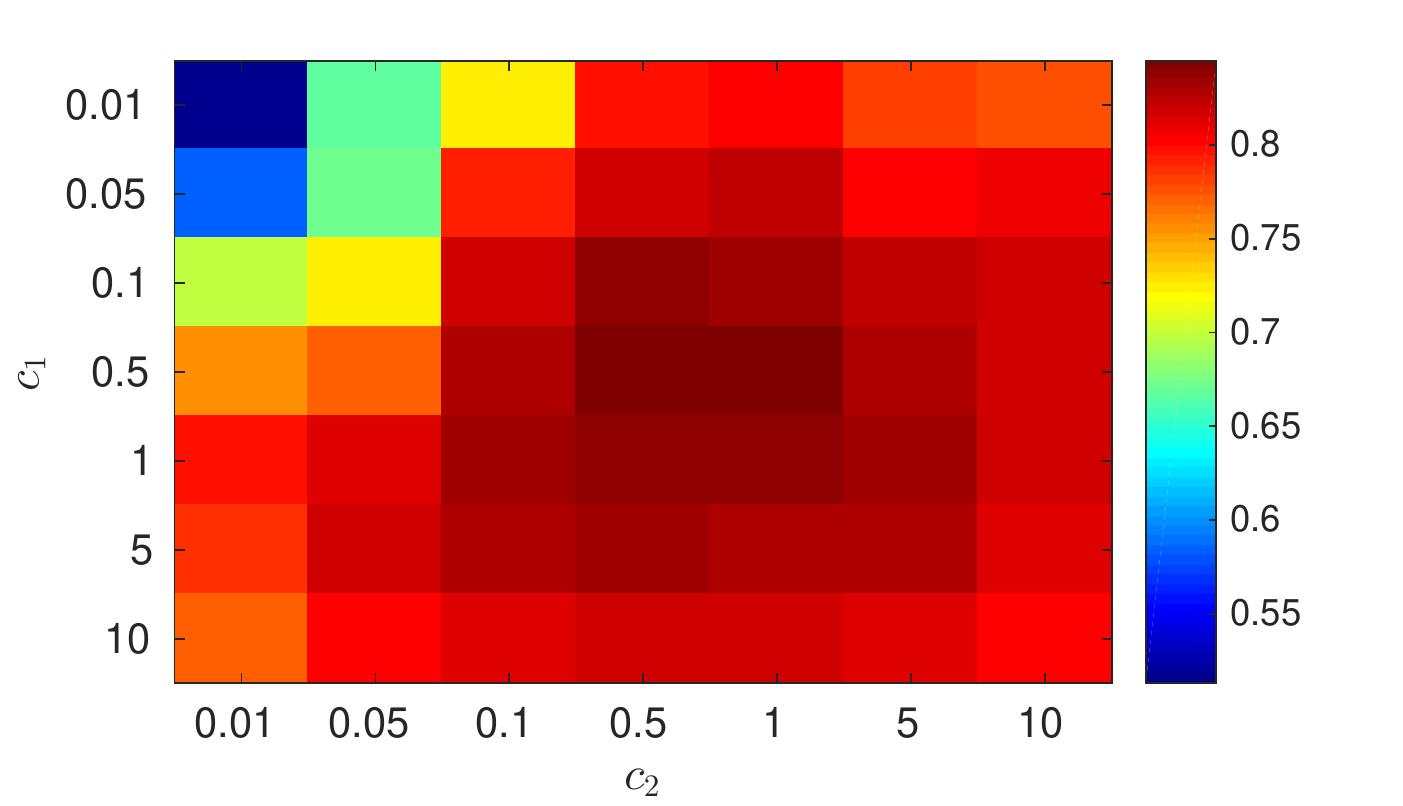}}
  \subfigure [DEAP dataset] {\includegraphics[height=0.85in,width=1.53in]{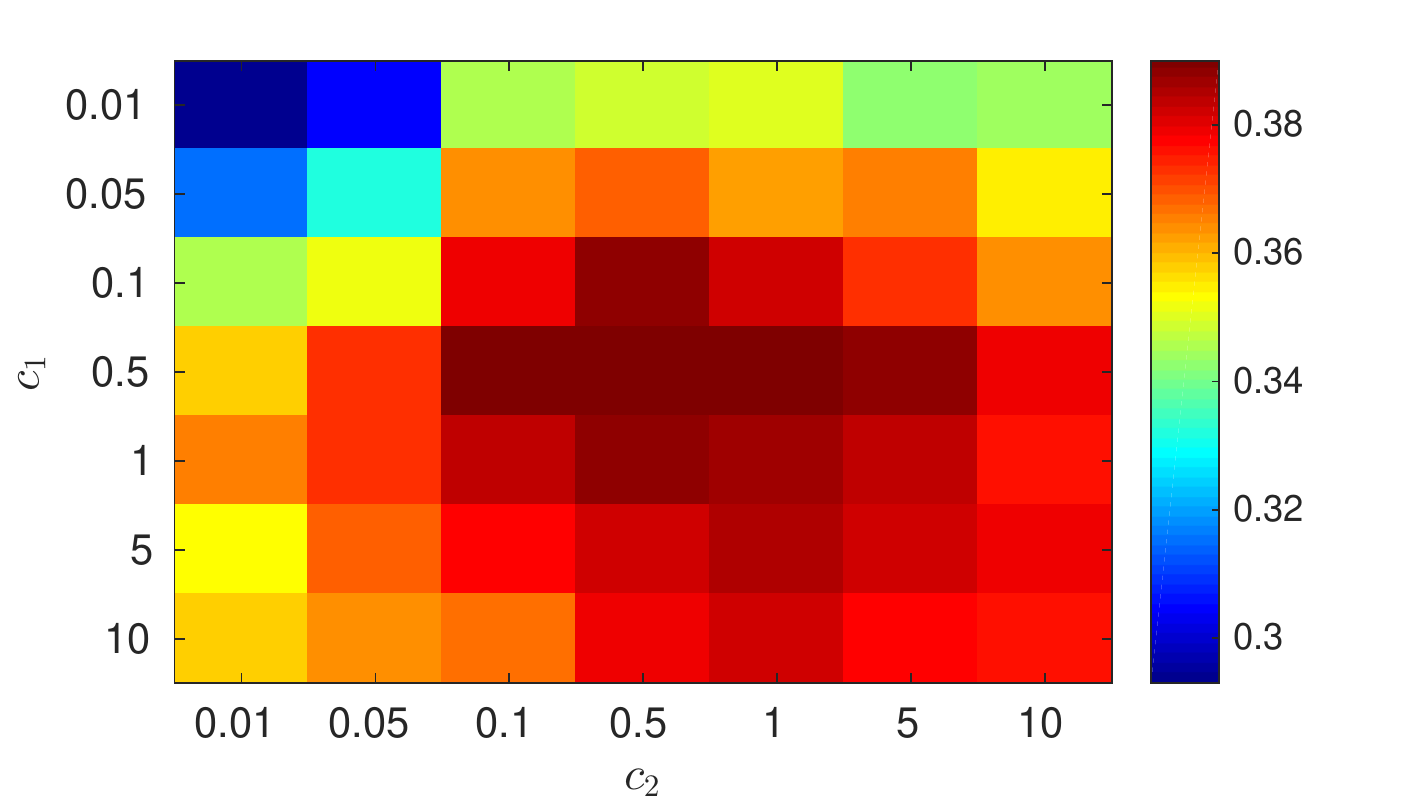}}
\caption{The impact of scaling constants $c_1$ and $c_2$.}
\label{fig:c1c2}
\end{figure}

Figure \ref{fig:c1c2} shows the classification accuracies of inductive SiMVAE with different scaling constants $c_1$ and $c_2$ on both datasets ($1\%$ labeled and 10\% missing data). From it, we can find that SiMVAE is not very sensitive to the values of $c_1$ and $c_2$. We choose the best $c_1$ and $c_2$ from \{0.1, 0.5, 1\} in the experiments.

\section{Conclusion}
We have proposed a novel semi-supervised multi-view deep generative framework for multi-modal emotion recognition with incomplete data. Under our framework, each modality of the emotional data is treated as one view, and the importance of each modality is inferred automatically by learning a non-uniformly weighted Gaussian mixture posterior approximation for the shared latent variable. The labeled-data-scarcity problem is naturally addressed within our framework through casting the semi-supervised classification problem as a specialized missing data imputation task. The incomplete-data problem is elegantly circumvented by treating the missing views as latent variables and integrating them out.
Compared with previous emotion recognition methods, our method is more robust and flexible.
Experimental results confirmed the superiorities of our framework over many state-of-the-art competitors.

\begin{acks}
This work was supported by National Natural Science Foundation of China (No. 91520202, 61602449), Beijing Municipal Science\&Technology Commission (Z181100008918010), Youth Innovation Promotion Association CAS and Strategic Priority Research Program of CAS.
\end{acks}

\bibliographystyle{ACM-Reference-Format}
\bibliography{changdedu}

\newpage

\centerline{\huge{\textbf{Supplementary Material}}}
\smallskip

In this document, we provide additional materials to supplement our main submission. In Section A, we provide further details on how we derive a second lower bound to variational lower bound. In Section B, we show Monte-Carlo estimators used to compute the gradients of the objective function.
\section*{Section A}
The Shannon entropy $\mathbb{E}_{q_{\phi}(\mathrm{\mathbf{z}}|\mathfrak{X},y)}[- \log q_{\phi}(\mathrm{\mathbf{z}}|\mathfrak{X},y)]$ is hard to compute analytically. In general, there is no closed-form expression for the entropy of a Mixture of Gaussians (MoG). Here we lower bound the entropy of MoG using Jensen's inequality:
\begin{align}
\nonumber  & \mathbb{E}_{q_{\phi}(\mathrm{\mathbf{z}}|\mathfrak{X},y)}[- \log q_{\phi}(\mathrm{\mathbf{z}}|\mathfrak{X},y)]  \\
\nonumber & = - \int q_{\phi}(\mathrm{\mathbf{z}}|\mathfrak{X},y) \log q_{\phi}(\mathrm{\mathbf{z}}|\mathfrak{X},y) \ d \mathrm{\mathbf{z}} \\
\nonumber  & = - \sum_{v=1}^2 \lambda^{(v)} \cdot \int \mathcal{N}\big(\mathrm{\mathbf{z}} | \bm{\mu}_{\phi^{(v)}},\ \bm{\Sigma}_{\phi^{(v)}} \big) \log \sum_{l=1}^2 \lambda^{(l)} \\
 \nonumber &  \hspace{5cm} \cdot \mathcal{N}\big(\mathrm{\mathbf{z}} | \bm{\mu}_{\phi^{(l)}},\ \bm{\Sigma}_{\phi^{(l)}} \big) \ d \mathrm{\mathbf{z}} \\
\nonumber  & \geq - \sum_{v=1}^2 \lambda^{(v)} \cdot \log \sum_{l=1}^2 \lambda^{(l)} \cdot \int \mathcal{N}\big(\mathrm{\mathbf{z}} | \bm{\mu}_{\phi^{(v)}},\ \bm{\Sigma}_{\phi^{(v)}} \big)  \\
 \nonumber &  \hspace{5cm} \cdot \mathcal{N}\big(\mathrm{\mathbf{z}} | \bm{\mu}_{\phi^{(l)}},\ \bm{\Sigma}_{\phi^{(l)}} \big) \ d \mathrm{\mathbf{z}} \\
\nonumber  & = - \sum_{v=1}^2 \lambda^{(v)} \cdot \log \sum_{l=1}^2 \lambda^{(l)} \cdot \mathcal{N}\big( \bm{\mu}_{\phi^{(v)}} | \bm{\mu}_{\phi^{(l)}},\ \bm{\Sigma}_{\phi^{(v)}} + \bm{\Sigma}_{\phi^{(l)}} \big)\\
\nonumber  & = - \sum_{v=1}^2 \lambda^{(v)} \cdot \log \bigg(\sum_{l=1}^2 \lambda^{(l)} \cdot \omega_{v, l}\bigg),
\end{align}
where we have used the fact that the convolution of two Gaussians is another Gaussian, and $\omega_{v, l} = \mathcal{N}\big( \bm{\mu}_{\phi^{(v)}} | \bm{\mu}_{\phi^{(l)}},\ \bm{\Sigma}_{\phi^{(v)}} + \bm{\Sigma}_{\phi^{(l)}} \big)$.
\section*{Section B}
$\mathbb{E}_{q_{\phi}(\mathrm{\mathbf{z}}|\mathfrak{X},y)}[\log p_{\theta^{(v)}}(\mathrm{\mathbf{x}}^{(v)}|y, \mathrm{\mathbf{z}})]$  can be rewritten, using the location-scale transformation for the Gaussian distribution, as:
\begin{align}
\label{reparameterization}
\nonumber  & \mathbb{E}_{q_{\phi}(\mathrm{\mathbf{z}}|\mathfrak{X},y)}[\log p_{\theta^{(v)}}(\mathrm{\mathbf{x}}^{(v)}|y, \mathrm{\mathbf{z}})]\\
\nonumber &= \sum_{l=1}^2 \lambda^{(l)} \mathbb{E}_{\mathcal{N}(\bm{\epsilon}^{(l)}|\mathrm{\mathbf{0}},\mathrm{\mathbf{I}})}\bigg[\log p_{\theta^{(v)}} (\mathrm{\mathbf{x}}^{(v)}|y, \bm{\mu}_{\phi^{(l)}} + \mathrm{\mathbf{R}}_{\phi^{(l)}} \bm{\epsilon}^{(l)} ) \bigg],
\end{align}
where $\mathrm{\mathbf{R}}_{\phi^{(l)}}\mathrm{\mathbf{R}}_{\phi^{(l)}}^\top = \bm{\Sigma}_{\phi^{(l)}}$ and  $l\in\{1, 2\}$.
While the expectations on the right hand side of the above equation still cannot be solved analytically, their gradients w.r.t. $\theta^{(v)}$, $\phi^{(l)}$ and $\lambda^{(l)}$ can be efficiently estimated using the following Monte-Carlo estimators
\begin{align}
\nonumber  &   \frac{\partial}{\partial \theta^{(v)}} \mathbb{E}_{q_{\phi}(\mathrm{\mathbf{z}}|\mathfrak{X},y)}[\log p_{\theta^{(v)}}(\mathrm{\mathbf{x}}^{(v)}|y, \mathrm{\mathbf{z}})] \\
\nonumber  &  \qquad = \sum_{l=1}^2  \lambda^{(l)} \mathbb{E}_{\mathcal{N}(\bm{\epsilon}^{(l)} |\mathrm{\mathbf{0}},\mathrm{\mathbf{I}})}\bigg[\frac{\partial }{\partial \theta^{(v)}} \log p_{\theta^{(v)}}\big (\mathrm{\mathbf{x}}^{(v)}|y, \mathrm{\mathbf{z}}^{(l)} \big )\bigg] \\
\nonumber  &  \qquad \approx \frac{\lambda^{(l)}}{T}  \sum_{t=1}^T \sum_{l=1}^2 \frac{\partial}{\partial \theta^{(v)}} \log p_{\theta^{(v)}} \big(\mathrm{\mathbf{x}}^{(v)}|y, \mathrm{\mathbf{z}}^{(l,t)} \big),  \qquad \qquad \qquad \qquad \qquad \qquad \ \ \ \quad
\end{align}
\vskip -0.1in
\begin{align}
\nonumber  &   \frac{\partial}{\partial \phi^{(l)}} \mathbb{E}_{q_{\phi}(\mathrm{\mathbf{z}}|\mathfrak{X},y)}[\log p_{\theta^{(v)}}(\mathrm{\mathbf{x}}^{(v)}|y, \mathrm{\mathbf{z}})] \\
\nonumber  & \qquad  = \lambda^{(l)} \frac{\partial}{\partial \phi^{(l)}} \mathbb{E}_{\mathcal{N}(\bm{\epsilon}^{(l)}|\mathrm{\mathbf{0}},\mathrm{\mathbf{I}})}\Big[\frac{\partial }{\partial \mathrm{\mathbf{z}}^{(l)} } \log p_{\theta^{(v)}}(\mathrm{\mathbf{x}}^{(v)}|y, \mathrm{\mathbf{z}}^{(l)}) \\
\nonumber &  \hspace{4.5cm} \cdot \Big( \frac{\partial \bm{\mu}_{\phi^{(l)}} }{\partial \phi^{(l)}} + \frac{\partial \mathrm{\mathbf{R}}_{\phi^{(l)}}}{\partial \phi^{(l)}} \bm{\epsilon}^{(l)} \Big) \Big] \\
\nonumber & \qquad  \approx \ \frac{\lambda^{(l)}}{T}\sum_{t=1}^T  \frac{\partial }{\partial \mathrm{\mathbf{z}}^{(l,t)} } \log p_{\theta^{(v)}}(\mathrm{\mathbf{x}}^{(v)}|y, \mathrm{\mathbf{z}}^{(l,t)})  \\
\nonumber &  \hspace{4.5cm} \cdot \Big( \frac{\partial \bm{\mu}_{\phi^{(l)}} }{\partial \bm{\phi}^{(l)}} + \frac{\partial \mathrm{\mathbf{R}}_{\phi^{(l)}}}{\partial \phi^{(l)}} \bm{\epsilon}^{(l,t)} \Big), \qquad\qquad
\end{align}
\vskip -0.1in
\begin{align}
\nonumber  &   \frac{\partial}{\partial \lambda^{(l)}} \mathbb{E}_{q_{\phi}(\mathrm{\mathbf{z}}|\mathfrak{X},y)}[\log p_{\theta^{(v)}}(\mathrm{\mathbf{x}}^{(v)}|y, \mathrm{\mathbf{z}})] \\
\nonumber  & \qquad  =  \mathbb{E}_{\mathcal{N}(\bm{\epsilon}^{(l)}|\mathrm{\mathbf{0}},\mathrm{\mathbf{I}})}\big[\log p_{\theta^{(v)}} (\mathrm{\mathbf{x}}^{(v)}|y, \mathrm{\mathbf{z}}^{(l)} ) \big]\\
\nonumber  &  \qquad \approx  \frac{1}{T}\sum_{t=1}^T  \log p_{\theta^{(v)}} \big(\mathrm{\mathbf{x}}^{(v)}|y, \mathrm{\mathbf{z}}^{(l,t)} \big),  \qquad\qquad\quad \qquad \qquad \qquad \qquad \qquad \qquad \ \ \ \quad
\end{align}
where $\mathrm{\mathbf{z}}^{(l)}$ is evaluated at $\mathrm{\mathbf{z}}^{(l)} = \bm{\mu}_{\phi^{(l)}} + \mathrm{\mathbf{R}}_{\phi^{(l)}} \bm{\epsilon}^{(l)}$ and $\mathrm{\mathbf{z}}^{(l,t)} = \bm{\mu}_{\phi^{(l)}} + \mathrm{\mathbf{R}}_{\phi^{(l)}} \bm{\epsilon}^{(l,t)}$ with $\bm{\epsilon}^{(l,t)} \sim \mathcal{N}(\mathrm{\mathbf{0}},\mathrm{\mathbf{I}})$. In practice, it suffices to use a small $T$ (e.g. $T = 1$) and then estimate the gradient using minibatches of data points. Though the above Monte-Carlo estimators could have large variances if a small $T$ is used, the experimental results show that it suffices to obtain good performance.
The same observation can be found in previous works \cite{sonderby2016ladder,maaloe2016auxiliary}. Furthermore, we use the same random numbers $\bm{\epsilon}^{(l,t)}$ for all estimators to have lower variances. The gradient w.r.t. $\varphi$ is omitted here, since it can be derived straightforwardly by using traditional reparameterization trick~\cite{kingma2014semi}.

\end{document}